\documentclass[twocolumn]{aastex631}

\usepackage{graphicx}	
\usepackage{amsmath}	
\usepackage{amssymb}	
\usepackage{xcolor}
\usepackage{subfigure}

\newcommand{\arcsecpoint}{\mbox{$''\!.$}}

\newcommand{\kms}{\mbox{km\,s$^{-1}$}}

\newcommand{\Msun}{\mbox{$M_\odot$}}

\newcommand{\hb}{\mbox{H$\beta$}}

\newcommand{\lya}{\mbox{Ly$\alpha$}}

\def\e0{\epsilon_0}
\def\HST{{HST}}

\def\Swift{{Swift}}

\def\hei{He\,{\sc i}}

\def\ciii{\ifmmode {\rm C}\,{\sc iii} \else C\,{\sc iii}\fi}
\def\civ{\ifmmode {\rm C}\,{\sc iv} \else C\,{\sc iv}\fi}

\def\nvii{N\,{\sc vii}}

\def\o5007{[O\,{\sc iii}]\,$\lambda5007$}

\def\ovii{O\,{\sc vii}}
\def\oviii{O\,{\sc viii}}

\def\siIV{Si\,{\sc iv}}

\def\pv{P\,{\sc v}}

\def\Chandra{{Chandra}}
\def\nustar{{NuSTAR}}
\def\nicer{{NICER}}

\def\relxillD{{\tt relxillD}}

\newcommand{\xmm}{{ XMM-Newton }}

\begin{document}
\title[AGN STORM 2. First results on Mrk 817]{AGN STORM 2: I. First results: A Change in the Weather of Mrk~817 }


\author[0000-0003-0172-0854]{Erin Kara}
\affiliation{MIT Kavli Institute for Astrophysics and Space Research, Massachusetts Institute of Technology, Cambridge, MA 02139, USA}

\author[0000-0002-4992-4664]{Missagh Mehdipour}
\affiliation{Space Telescope Science Institute, 3700 San Martin Drive, Baltimore, MD 21218, USA}

\author[0000-0002-2180-8266]{Gerard A.\ Kriss}
\affiliation{Space Telescope Science Institute, 3700 San Martin Drive, Baltimore, MD 21218, USA}

\author[0000-0002-8294-9281]{Edward M.\ Cackett}
\affiliation{Department of Physics and Astronomy, Wayne State University, 666 W.\ Hancock St, Detroit, MI, 48201, USA}

\author[0000-0003-2991-4618]{Nahum Arav}
\affiliation{Department of Physics, Virginia Tech, Blacksburg, VA 24061, USA}

\author[0000-0002-3026-0562]{Aaron J.\ Barth}
\affiliation{Department of Physics and Astronomy, 4129 Frederick Reines Hall, University of California, Irvine, CA, 92697-4575, USA}

\author{Doyee Byun}
\affiliation{Department of Physics, Virginia Tech, Blacksburg, VA 24061, USA}

\author[0000-0002-1207-0909]{Michael S.\ Brotherton}
\affiliation{Department of Physics and Astronomy, University of Wyoming, Laramie, WY 82071, USA}

\author[0000-0003-3242-7052]{Gisella De~Rosa}
\affiliation{Space Telescope Science Institute, 3700 San Martin Drive, Baltimore, MD 21218, USA}

\author[0000-0001-9092-8619]{Jonathan Gelbord}
\affiliation{Spectral Sciences Inc., 4 Fourth Ave., Burlington, MA 01803, USA}

\author[0000-0002-6733-5556]{Juan V.\ Hern\'{a}ndez Santisteban}
\affiliation{SUPA School of Physics and Astronomy, North Haugh, St.~Andrews, KY16~9SS, Scotland, UK}

\author{Chen Hu}
\affiliation{Key Laboratory for Particle Astrophysics, Institute of High Energy Physics, Chinese Academy of Sciences, 19B Yuquan Road,\\ Beijing 100049, People's Republic of China}

\author[0000-0001-5540-2822]{Jelle Kaastra}
\affiliation{SRON Netherlands Institute for Space Research, Sorbonnelaan 2, 3584 CA Utrecht, The Netherlands}
\affiliation{Leiden Observatory, Leiden University, PO Box 9513, 2300 RA Leiden, The Netherlands}

\author{Hermine Landt}
\affiliation{Centre for Extragalactic Astronomy, Department of Physics, Durham University, South Road, Durham DH1 3LE, UK}

\author[0000-0001-5841-9179]{Yan-Rong Li}
\affiliation{Key Laboratory for Particle Astrophysics, Institute of High Energy Physics, Chinese Academy of Sciences, 19B Yuquan Road,\\ Beijing 100049, People's Republic of China}

\author[0000-0001-8475-8027]{Jake A.\ Miller}
\affiliation{Department of Physics and Astronomy, Wayne State University, 666 W.\ Hancock St, Detroit, MI, 48201, USA}

\author[0000-0001-5639-5484]{John Montano}
\affiliation{Department of Physics and Astronomy, 4129 Frederick Reines Hall, University of California, Irvine, CA, 92697-4575, USA}

\author[0000-0003-1183-1574]{Ethan Partington}
\affiliation{Department of Physics and Astronomy, Wayne State University, 666 W.\ Hancock St, Detroit, MI, 48201, USA}


\author[0000-0003-0487-1105]{Jes\'{u}s Aceituno}  
\affiliation{Centro Astronomico Hispano Alem\'{a}n, Sierra de los filabres sn, E-04550 G\'{e}rgal, Almer\'{\i}a, Spain}
\affiliation{Instituto de Astrof\'{\i}sica de Andaluc\'{\i}a (CSIC), Glorieta de la astronom\'{\i}a sn, E-18008 Granada, Spain}

\author{Jin-Ming Bai}
\affiliation{Yunnan Observatories, The Chinese Academy of Sciences, Kunming 650011, People's Republic of China}

\author{Dongwei Bao} 
\affiliation{Key Laboratory for Particle Astrophysics, Institute of High Energy Physics, Chinese Academy of Sciences, 19B Yuquan Road,\\ Beijing 100049, People's Republic of China}
\affil{School of Astronomy and Space Science, University of Chinese Academy of Sciences, 19A Yuquan Road, Beijing 100049, China}

\author[0000-0002-2816-5398]{Misty C.\ Bentz}
\affiliation{Department of Physics and Astronomy, Georgia State University, 25 Park Place, Suite 605, Atlanta, GA 30303, USA}

\author[0000-0001-5955-2502]{Thomas G.\ Brink}
\affiliation{Department of Astronomy, University of California, Berkeley, CA 94720-3411, USA}


\author[0000-0002-4830-7787]{Doron Chelouche}
\affiliation{Department of Physics, Faculty of Natural Sciences, University of Haifa, Haifa 3498838, Israel}
\affiliation{Haifa Research Center for Theoretical Physics and Astrophysics, University of Haifa, Haifa 3498838, Israel }

\author{Yong-Jie Chen}
\affiliation{Key Laboratory for Particle Astrophysics, Institute of High Energy Physics, Chinese Academy of Sciences, 19B Yuquan Road,\\ Beijing 100049, People's Republic of China}
\affiliation{School of Astronomy and Space Science, University of Chinese Academy of Sciences, 19A Yuquan Road, Beijing 100049, China}

\author[0000-0001-9931-8681]{Elena Dalla Bont\`{a}}
\affiliation{Dipartimento di Fisica e Astronomia ``G.\  Galilei,'' Universit\'{a} di Padova, Vicolo dell'Osservatorio 3, I-35122 Padova, Italy}
\affiliation{INAF-Osservatorio Astronomico di Padova, Vicolo dell'Osservatorio 5 I-35122, Padova, Italy}

\author[0000-0002-0964-7500]{Maryam Dehghanian}
\affiliation{Department of Physics and Astronomy, The University of Kentucky, Lexington, KY 40506, USA}

\author[0000-0002-5830-3544]{Pu Du} 
\affiliation{Key Laboratory for Particle Astrophysics, Institute of High Energy Physics, Chinese Academy of Sciences, 19B Yuquan Road,\\ Beijing 100049, People's Republic of China}
\author[0000-0001-8598-1482]{Rick Edelson} 
\affiliation{Eureka Scientific Inc., 2452 Delmer St. Suite 100, Oakland, CA 94602, USA}
\author[0000-0003-4503-6333]{Gary J.\ Ferland}
\affiliation{Department of Physics and Astronomy, The University of Kentucky, Lexington, KY 40506, USA}

\author[0000-0002-8224-1128]{Laura Ferrarese}
\affiliation{NRC Herzberg Astronomy and Astrophysics Research Centre, 5071 West Saanich Road, Victoria, BC, V9E 2E7, Canada}

\author{Carina Fian}
\affiliation{School of Physics and Astronomy and Wise observatory, Tel Aviv University, Tel Aviv 6997801, Israel}
\affiliation{Haifa Research Center for Theoretical Physics and Astrophysics, University of Haifa, Haifa 3498838, Israel }

\author[0000-0003-3460-0103]{Alexei V.\ Filippenko}
\affiliation{Department of Astronomy, University of California, Berkeley, CA 94720-3411, USA}
\affiliation{Miller Institute for Basic Research in Science, University of California, Berkeley, CA  94720, USA}

\author[0000-0002-3365-8875]{Travis Fischer}
\affiliation{AURA for ESA, Space Telescope Science Institute, Baltimore, MD, USA, 3700 San Martin Drive, Baltimore, MD 21218, USA}


\author[0000-0002-2908-7360]{Michael R.\ Goad}
\affiliation{School of Physics and Astronomy, University of Leicester, University Road, Leicester, LE1 7RH, UK}

\author[0000-0002-9280-1184]{Diego H.\ Gonz\'{a}lez Buitrago}
\affiliation{Instituto de Astronom\'{\i}a, Universidad Nacional Aut\'{o}noma de M\'{e}xico, Km 103 Carretera Tijuana-Ensenada, 22860 Ensenada B.C., M\'{e}xico}

\author[0000-0002-8990-2101]{Varoujan Gorjian}
\affiliation{Jet Propulsion Laboratory, M/S 169-327, 4800 Oak Grove Drive, Pasadena, CA 91109, USA}

\author[0000-0001-9920-6057]{Catherine J.\ Grier}
\affiliation{Steward Observatory, University of Arizona, 933 North Cherry Avenue, Tucson, AZ 85721, USA}

\author{Wei-Jian Guo}
\affiliation{Key Laboratory for Particle Astrophysics, Institute of High Energy Physics, Chinese Academy of Sciences, 19B Yuquan Road,\\ Beijing 100049, People's Republic of China}
\affiliation{School of Astronomy and Space Science, University of Chinese Academy of Sciences, 19A Yuquan Road, Beijing 100049, China}

\author[0000-0002-1763-5825]{Patrick B.\ Hall}
\affiliation{Department of Physics and Astronomy, York University, Toronto, ON M3J 1P3, Canada}

\author[0000-0002-0957-7151]{Y. Homayouni}
\affiliation{Space Telescope Science Institute, 3700 San Martin Drive, Baltimore, MD 21218, USA}

\author[0000-0003-1728-0304]{Keith Horne}
\affiliation{SUPA School of Physics and Astronomy, North Haugh, St.~Andrews, KY16~9SS, Scotland, UK}


\author[0000-0002-1134-4015]{Dragana Ili\'{c}}
\affiliation{Department of Astronomy, Faculty of Mathematics, University of
Belgrade, Studentski trg 16,11000 Belgrade, Serbia}
\affiliation{Humboldt Research Fellow, Hamburger Sternwarte, Universit{\"a}t
Hamburg, Gojenbergsweg 112, 21029 Hamburg, Germany}


\author{Bo-Wei Jiang}
\affiliation{Key Laboratory for Particle Astrophysics, Institute of High Energy Physics, Chinese Academy of Sciences, 19B Yuquan Road,\\ Beijing 100049, People's Republic of China}
\affiliation{School of Astronomy and Space Science, University of Chinese Academy of Sciences, 19A Yuquan Road, Beijing 100049, China}

\author[0000-0003-0634-8449]{Michael D.\ Joner}
\affiliation{Department of Physics and Astronomy, N283 ESC, Brigham Young University, Provo, UT 84602, USA}


\author[0000-0002-9925-534X]{Shai Kaspi}
\affiliation{School of Physics and Astronomy and Wise observatory, Tel Aviv University, Tel Aviv 6997801, Israel}

\author[0000-0001-6017-2961]{Christopher S.\ Kochanek}
\affiliation{Department of Astronomy, The Ohio State University, 140 W.\ 18th Ave., Columbus, OH 43210, USA}
\affiliation{Center for Cosmology and AstroParticle Physics, The Ohio State University, 191 West Woodruff Ave., Columbus, OH 43210, USA}

\author[0000-0003-0944-1008]{Kirk T.\ Korista}
\affiliation{Department of Physics, Western Michigan University, 1120 Everett Tower, Kalamazoo, MI 49008-5252, USA}

\author[0000-0001-8638-3687]{Daniel \ Kynoch}
\affiliation{Astronomical Institute of the Czech Academy of Sciences, Bo\v{c}n\'{i} II 1401/1a, CZ-14100 Prague, Czechia}


\author{Sha-Sha Li}
\affiliation{Key Laboratory for Particle Astrophysics, Institute of High Energy Physics, Chinese Academy of Sciences, 19B Yuquan Road,\\ Beijing 100049, People's Republic of China}
\affiliation{School of Astronomy and Space Science, University of Chinese Academy of Sciences, 19A Yuquan Road, Beijing 100049, China}

\author{Jun-Rong Liu}
\affiliation{Key Laboratory for Particle Astrophysics, Institute of High Energy Physics, Chinese Academy of Sciences, 19B Yuquan Road,\\ Beijing 100049, People's Republic of China}
\affiliation{School of Astronomy and Space Science, University of Chinese Academy of Sciences, 19A Yuquan Road, Beijing 100049, China}

\author{Ian M.\ M$^{\rm c}$Hardy}
\affiliation{School of Physics and Astronomy, University of Southampton, Highfield, Southampton SO17 1BJ, UK}

\author{Jacob N.\ McLane}
\affiliation{Department of Physics and Astronomy, University of Wyoming, Laramie, WY 82071, USA}

\author{Jake A. J.\ Mitchell}
\affiliation{Centre for Extragalactic Astronomy, Department of Physics, Durham University, South Road, Durham DH1 3LE, UK}
\author[0000-0002-6766-0260]{Hagai Netzer}
\affiliation{School of Physics and Astronomy, Tel Aviv University, Tel Aviv 69978, Israel}


\author{Kianna A.\ Olson}
\affiliation{Department of Physics and Astronomy, University of Wyoming, Laramie, WY 82071, USA}


\author[0000-0003-1435-3053]{Richard W.\ Pogge}
\affiliation{Department of Astronomy, The Ohio State University, 140 W.\ 18th Ave., Columbus, OH 43210, USA}
\affiliation{Center for Cosmology and AstroParticle Physics, The Ohio State University, 191 West Woodruff Ave., Columbus, OH 43210, USA}

\author[0000-0003-2398-7664]{Luka \u{C}.\ Popovi\'{c}}
\affiliation{Astronomical Observatory, Volgina 7, 11060 Belgrade, Serbia}
\affiliation{Department of Astronomy, Faculty of Mathematics, University of Belgrade, Studentski trg 16,11000 Belgrade, Serbia}

\author[0000-0002-6336-5125]{Daniel Proga}
\affiliation{Department of Physics \& Astronomy, 
University of Nevada, Las Vegas 
4505 S.\ Maryland Pkwy, 
Las Vegas, NV,
89154-4002, USA}



\author[0000-0003-1772-0023]{Thaisa Storchi-Bergmann}
\affiliation{Departamento de Astronomia - IF, Universidade Federal do Rio Grande do Sul, CP 150501, 91501-970 Porto Alegre, RS, Brazil}

\author[0000-0002-4930-0093]{Erika Strasburger}
\affiliation{Department of Astronomy, University of California, Berkeley, CA 94720-3411, USA}

\author[0000-0002-8460-0390]{Tommaso Treu}\thanks{Packard Fellow}
\affiliation{Department of Physics and Astronomy, University of California, Los Angeles, CA 90095, USA}


\author[0000-0001-9191-9837]{Marianne Vestergaard}
\affiliation{Steward Observatory, University of Arizona, 933 North Cherry Avenue, Tucson, AZ 85721, USA}
\affiliation{DARK, The Niels Bohr Institute, University of Copenhagen, Jagtvej 128, DK-2200 Copenhagen, Denmark}

\author[0000-0001-9449-9268]{Jian-Min Wang}
\affiliation{Key Laboratory for Particle Astrophysics, Institute of High Energy Physics, Chinese Academy of Sciences, 19B Yuquan Road,\\ Beijing 100049, People's Republic of China}

\affiliation{School of Astronomy and Space Sciences, University of Chinese Academy of Sciences, 19A Yuquan Road, Beijing 100049, People's Republic of China}
\affiliation{National Astronomical Observatories of China, 20A Datun Road, Beijing 100020, People's Republic of China}

\author[0000-0003-1810-0889]{Martin J.\ Ward}
\affiliation{Centre for Extragalactic Astronomy, Department of Physics, Durham University, South Road, Durham DH1 3LE, UK}

\author[0000-0002-5205-9472]{Tim Waters}
\affiliation{Department of Physics \& Astronomy, 
University of Nevada, Las Vegas 
4505 S. Maryland Pkwy, 
Las Vegas, NV, 89154-4002, USA}

\author[0000-0002-4645-6578]{Peter R.\ Williams}
\affiliation{Department of Physics and Astronomy, University of California, Los Angeles, CA 90095, USA}


\author{Sen Yang}
\affiliation{Key Laboratory for Particle Astrophysics, Institute of High Energy Physics, Chinese Academy of Sciences, 19B Yuquan Road,\\ Beijing 100049, People's Republic of China}
\affiliation{School of Astronomy and Space Science, University of Chinese Academy of Sciences, 19A Yuquan Road, Beijing 100049, China}

\author{Zhu-Heng Yao}
\affiliation{Key Laboratory for Particle Astrophysics, Institute of High Energy Physics, Chinese Academy of Sciences, 19B Yuquan Road,\\ Beijing 100049, People's Republic of China}
\affiliation{School of Astronomy and Space Science, University of Chinese Academy of Sciences, 19A Yuquan Road, Beijing 100049, China}


\author{Theodora E.\ Zastrocky}
\affiliation{Department of Physics and Astronomy, University of Wyoming, Laramie, WY 82071, USA}

\author{Shuo Zhai}
\affiliation{Key Laboratory for Particle Astrophysics, Institute of High Energy Physics, Chinese Academy of Sciences, 19B Yuquan Road,\\ Beijing 100049, People's Republic of China}
\affiliation{School of Astronomy and Space Science, University of Chinese Academy of Sciences, 19A Yuquan Road, Beijing 100049, China}

\author[0000-0001-6966-6925]{Ying Zu}
\affiliation{Department of Astronomy, School of Physics and Astronomy, Shanghai Jiao Tong University, Shanghai 200240, China}

\begin{abstract}

We present the first results from the ongoing, intensive, multi-wavelength monitoring program of the luminous Seyfert 1 galaxy Mrk~817. While this AGN was, in part, selected for its historically unobscured nature, we discovered that the X-ray spectrum is highly absorbed, and there are new blueshifted, broad and narrow UV absorption lines, which suggest that a dust-free, ionized obscurer located at the inner broad line region partially covers the central source. Despite the obscuration, we measure UV and optical continuum reverberation lags consistent with a centrally illuminated Shakura-Sunyaev thin accretion disk, and measure reverberation lags associated with the optical broad line region, as expected. However, in the first 55~days of the campaign, when the obscuration was becoming most extreme, we observe a de-coupling of the UV continuum and the UV broad emission line variability. The correlation recovers in the next 42 days of the campaign, as Mrk~817 enters a less obscured state. The short \civ~and \lya\ lags suggest that the accretion disk extends beyond the UV broad line region.

\end{abstract}


\section{Introduction}

For nearly two decades, quasars have been regarded as critical elements in our
understanding of galaxy evolution. Without energetic quasar
outflows in the forms of radiation, winds, or radio jet plasma, galaxy
evolution models overpredict the luminosities of galaxies at the
bright end of the luminosity function \citep{croton06,Fabian2012}. Quasar outflows may provide feedback that can heat
or remove the interstellar medium of the host galaxy. This can shut
down star formation and terminate the gas flow to the black hole,
thus freezing the total stellar luminosity and black hole mass and
leaving a quiescent black hole at the center of a dead
(non-star-forming) galaxy (e.g. \citealt{hopkins08}). The otherwise baffling correlations between
the mass of the central supermassive black hole and gross observable
properties of the host galaxy (e.g., bulge velocity dispersion or
luminosity) are naturally explained by this feedback \citep{ferrarese00, gebhardt00,yu02}. However, direct observational evidence of quasar feedback has
remained elusive.

Among the processes that are poorly understood are the gas flows 
and their origin near the
central black hole. There is ample evidence of large-scale, high-velocity
outflows from blueshifted X-ray absorption lines (e.g., \citealt{pounds03,tombesi10, parker17}) and UV spectra (e.g., \citealt{kriss18}) -- however, the information gleaned from this is only
along the line of sight, only in the resonance lines, and is difficult to convert into mass flows because of the unknown covering factor and unknown physical distance of the absorber to the black hole.
With the rare exceptions of recent interferometric observations of AGN including 3C273 \citep{GRAVITY2018} and 
NGC\,3783 \citep{GRAVITY2021},
the centi-parsec scales of AGN in the UV/optical
are too small to be directly resolvable. Resolving micro-parsec scales of the X-ray emitting region is even less attainable, except for the notable exception of the sub-mm image of the low-luminosity AGN in M87 \citep{EHT2019}. 
Consequently the most powerful tool available to probe
active nuclei is reverberation mapping (RM), which substitutes time
resolution for spatial resolution
\citep{Blandford1982}. 

The reverberation mapping technique was pioneered in the optical, where the location and motion of gas flows that constitute the ``broad-line region'' (BLR) are constrained by observing the time-delayed response of broad emission lines
to the variable continuum flux from the black hole accretion disk (e.g. \citealt{Peterson1993,bentz09}). In recent years, higher cadence observing campaigns at shorter wavelengths has allowed for reverberation mapping of the continuum-emitting accretion disk in the UV/optical (e.g. \citealt{Edelson2015, cackett18}), and even within the inner tens of gravitational radii with X-ray reverberation (e.g. \citealt{uttley14}). 

The UV wavelength regime is particularly important for understanding the inner centi-parsec gas flows in AGN using RM
because of the presence of
two of the strongest and  most important diagnostic emission lines, \civ\,$\lambda\lambda1548,$ 1551, and \lya\,$\lambda1215$. Moreover, the UV
continuum arises primarily in the inner accretion disk and is a better proxy than the optical continuum for the unobservable ionizing continuum that drives the line variations.

Several pioneering AGN UV spectral monitoring programs were undertaken
with the International Ultraviolet Explorer (IUE) and 
the Hubble Space Telescope (HST)
in the late 1980s and 1990s.  These early campaigns led to an
understanding of the practicalities of RM that informed more intensive campaigns, intended to
determine not only sizes of the line-emitting regions, but also their
kinematics \citep[e.g.,][]{Horne2004}. 
While ground-based optical campaigns designed for velocity-dependent RM (2D RM) have achieved great success, there has only been one 2D RM program in the UV. This study of NGC~5548
is known as the AGN Space Telescope and Optical Reverberation Mapping program (AGN STORM; e.g. \citealt{DeRosa2015,Kriss2019}). Here we present the first results from a second such
program, this time on Mrk~817. We refer to the new campaign as AGN STORM 2, and henceforth refer to the NGC 5548 campaign as AGN STORM 1.

The AGN STORM 1 program was a multiwavelength spectroscopic and photometric 
monitoring campaign that was undertaken in the first half of 2014 (\HST\ Cycle 21),
anchored by nearly daily observations of NGC 5548 over six months with the \HST\ Cosmic Origins Spectrograph (COS).
The \HST\ observations were supplemented by high-cadence (approximately twice daily)
X-ray and near-UV observations with \Swift\ \citep{Edelson2015}, as well as 
intensive ground-based optical photometric \citep{Fausnaugh2016} and
spectroscopic \citep{Pei2017} monitoring, plus four observations with
\Chandra\ \citep{Mathur2017}. 

\begin{figure*}[!tbp]
  \centering
  \includegraphics[width=\textwidth]{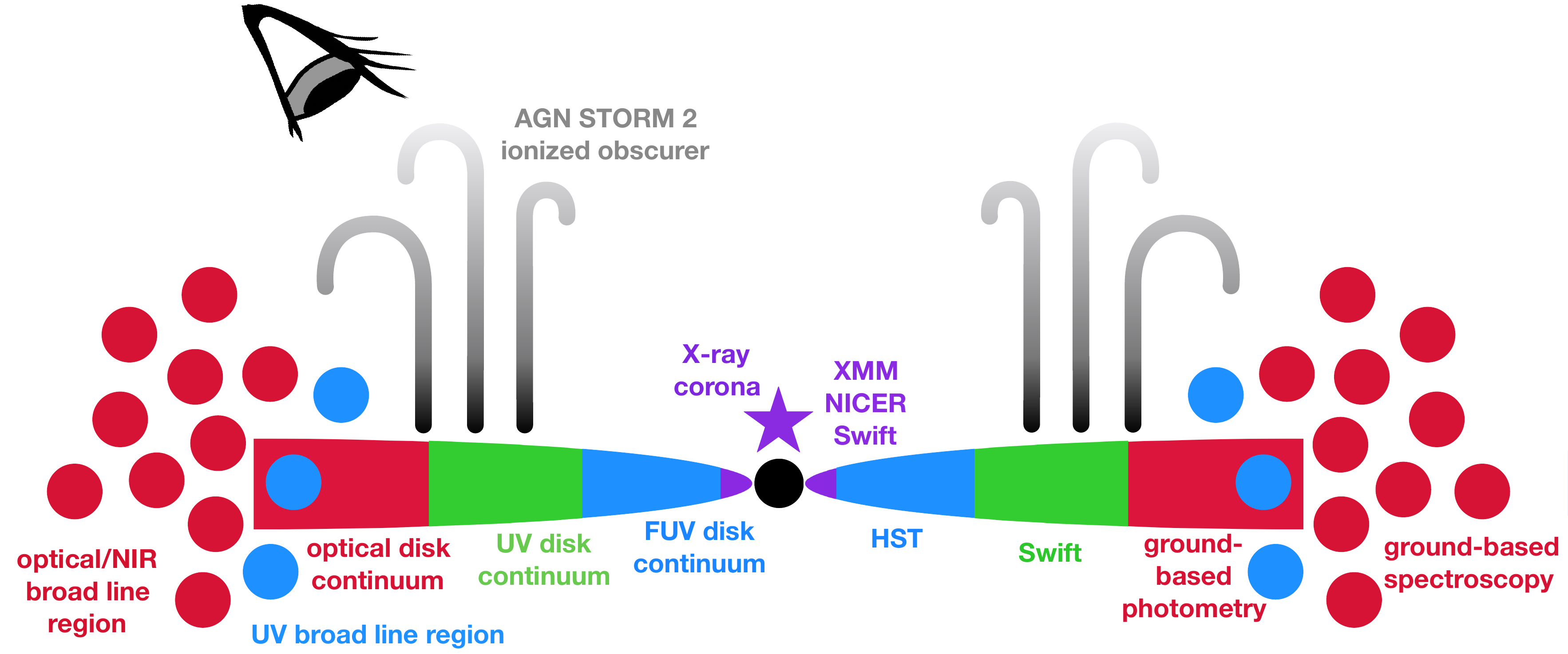}
  \caption{Schematic overview of the AGN~STORM~2 campaign, highlighting the telescopes involved thus far, and what regions they probe. The color schemes correspond to different telescopes/wavelength regimes, and will be used throughout this paper and future papers. The most unexpected result is the presence of a new dust-free ionized wind, as evidenced by the presence of new narrow and broad UV absorption lines (Fig~\ref{fig_cosfull}), and a significant depression of soft X-rays relative to earlier observations (Fig.~\ref{fig:x-ray}). Spectral decomposition (Section~\ref{sec:uv_results}, Section~\ref{sec:xray_results}) and photoionization modelling (Section~\ref{sec:photoionization}) place the new absorber at the inner broad line region. Despite this unexpected wind, the optical/UV continuum reverberation lags appear to behave as expected for reprocessing in a standard thin disk (Section~\ref{sec:continuum_rm} and Fig.~\ref{fig:contlags}). The broad line region reverberation is more complex, as the \hb\ lags behave as expected, but the UV broad emission lines are uncorrelated with the continuum for the first $\sim 60$~days, likely due to interference from the obscurer (Section~\ref{sec:blr_rm} and Fig.~\ref{fig:blrlags}).}
  \label{schematic}
\end{figure*}

AGN STORM 1 yielded a number of surprising results. The single most important lesson from this campaign is that intensive multi-wavelength monitoring enables us to decipher the relationships between different spectral properties including the presence (or lack) of reverberation, and the role of obscuration in the reprocessing of radiation. Among the interesting and unexpected results from AGN
STORM~1 are the following:
\begin{enumerate}
\item Longer-wavelength continuum variations follow shorter-wavelength
continuum variations from the UV through the NIR
in a pattern largely consistent with the temperature
gradient expected from a standard Shakura--Sunyaev thin accretion disk
\citep{Shakura1973},
but suggesting a disk that is three times larger
\citep{Edelson2015, Fausnaugh2016, Starkey2017}.
However, the X-ray variations
and those at longer wavelengths are not simply related
\citep[e.g.,][]{Gardner2017, edelson19}.
\item The emission-line lags were much smaller than expected from the well-established relationship between the AGN luminosity and the
time-delayed response of the broad emission lines. Given the
luminosity of NGC 5548 and its past behavior, the expected \hb\ lag was $\sim 20$\,days,
but the observed delay was only $\sim 6$\,days \citep{Pei2017}.
\item The time lag between the UV and optical continuum variations 
is the same as the time lag between the UV continuum variations and the
most rapidly responding emission lines (\ion{He}{2} $\lambda1640$
and \ion{He}{2} $\lambda4686$, $\sim 2$\,days; \citealt{fausnaugh16}).  
\item During part of the AGN STORM 1 program, all of the emission lines
and the high-ionization absorption lines apparently decoupled from
the continuum variations (the so-called ``BLR holiday'' of \citealt{Goad2016}). 
\item The ``velocity-delay maps,'' i.e., the projection
of the BLR into the two observables of time delay and line-of-sight
velocity, suggest the presence of an inclined disk, but
the response of the far side of the disk is weaker than expected
\citep{Horne2021}. Direct modeling of the spectra yields similar results \citep{Williams2020}.
\end{enumerate}
The emission-line and absorption-line behaviors can be explained by an ``obscurer''
\citep[cf.,][]{Kaastra2014} that is located in the inner BLR and associated with the broad absorption seen in the short wavelength wings of the
broad resonance emission lines. The properties of the obscurer are consistent with a disk wind that is launched from the inner BLR, and whose absorption alters the SED seen by more distant emission components. Instabilities in the density of the base of the wind cause variations in the line of sight covering factor \citep{Dehghanian2019,Dehghanian2020,Dehghanian2021}.  The existence of an obscurer in NGC 5548 underscored the importance of contemporaneous X-ray observations. This enables us to determine the SED that is incident upon the obscurer, and the SED that is filtered by the obscurer and reaches the BLR. 

In this paper, we present the first results from a second intensive multiwavelength campaign intended to study gas flows, as manifested in both emission and absorption lines, in the vicinity of a supermassive black hole. The target of this new campaign is Mrk 817 (PG 1434+590). Mrk 817 was selected for this campaign based on a number of considerations, both scientific and practical:
\begin{enumerate}
    \item There is no historical evidence
    for the type of broad UV absorption lines that complicated the interpretation of the NGC 5548 data. We believed that the sightline to the BLR and accretion disk would be clear and unobscured.
    \item Mrk 817 affords an opportunity to explore a different part of AGN parameter space; its mass ($M_{\rm BH} \approx 3.85 \times 10^7\,\Msun$) is similar to that of NGC 5548
    but it is more luminous and so has a higher Eddington ratio ($L/L_{\mathrm{Edd}}\sim 0.2$ for Mrk~817, and $L/L_{\mathrm{Edd}}\sim 0.03$ for NGC~5548).
    \item The Galactic foreground extinction is low, $E(B-V) \approx 0.02$\,mag, and Mrk 817 is  close enough to the north ecliptic pole that it can be observed throughout the year by \HST\ and other satellites and moderate-to-high latitude ground-based observatories.
    \end{enumerate}

In this paper, we discuss some early results and findings from the first third of the campaign.
These early results do not incorporate planned improvements to the COS and STIS calibrations that are also part of our ongoing program. Final results upon completion of the program may differ slightly from those presented here.

The biggest surprise in our observations was that there are now both broad and narrow absorption features in the UV spectra, and the X-rays show that the ionizing continuum is heavily obscured (see schematic in Fig~\ref{schematic}). This appears to have induced a ``broad line holiday'' as the broad UV emission lines in the early part of the campaign do not respond coherently to UV continuum variations. Remarkably, however, the broad \hb\ emission line does appear to lag behind the continuum with the time delay expected from the radius-luminosity relation \citep{bentz13}.

The paper is organized as follows: In Section~\ref{sec:obs}, we describe the data from the space and ground-based telescopes involved in the campaign. In Section~\ref{sec:dec2020}, we present the detailed spectral modelling of the long HST STIS and COS observation (Visit 3N) and a long \xmm\ stare, both taken on 2020-12-18. We focus on characterizing the new ionized obscurer, and in Section~\ref{sec:evolution}, we put the results into the context of the overall AGN~STORM~2 campaign thus far (97~days, from 2020-11-11 to 2021-03-01; HJD 2459177-2459274). Next, in Section~\ref{sec:photoionization}, we  present photoionization models for the narrow and broad UV absorption lines, given the ionizing broadband SED from the 2020-12-18 observations. Finally in \ref{sec:lags}, we present preliminary UV/optical continuum and broad line region reverberation mapping.  

\section{Observations and Data Reduction}

\begin{figure*}[!tbp]
  \centering
   \resizebox{1.1\hsize}{!}{\hspace*{-40pt}\includegraphics[angle=0, width=1.20\textwidth]{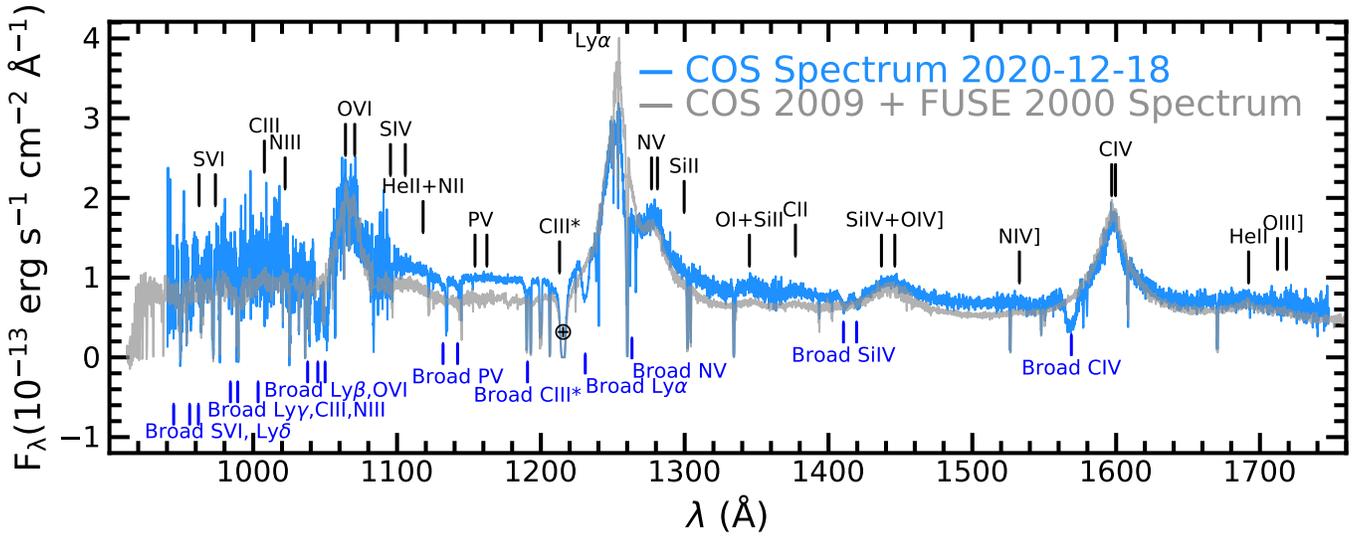}}
  \vskip -220pt
  \caption{The COS spectrum of Mrk 817 from 2020-12-18 (blue), compared to archival spectra (gray) from COS (in 2009; \citealt{Winter11})
and FUSE (in 2000; \citealt{kriss01}).
The COS data are binned by 8 pixels ($\sim$1 resolution element).
Positions of typical AGN emission features are labeled above the spectrum,
and new broad absorption features detected in 2020 are labeled in blue.
Geocoronal emission is indicated with an Earth symbol in the center of the
Milky Way Ly$\alpha$ absorption trough, and has been removed.}
  \label{fig_cosfull}
\end{figure*}

\begin{figure*}
  \centering
  \includegraphics[width=0.66\textwidth]{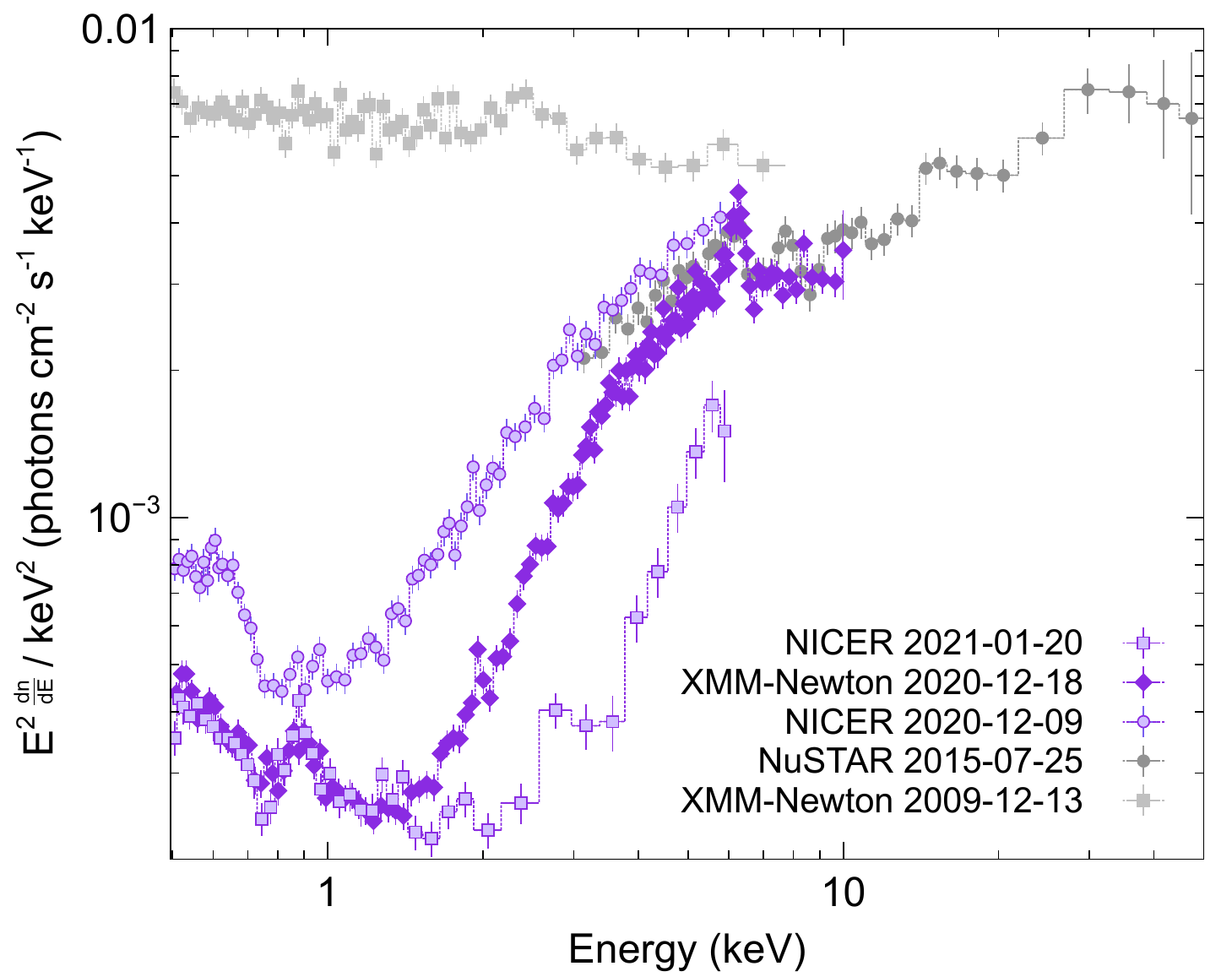}
  \caption{X-ray counts multiplied by an $E^{2}$ powerlaw, illustrating the change in spectral shape between earlier observations from \xmm\ in 2009 and \nustar\ in 2015 (gray), and the new observations (XMM-Newton in purple diamonds and \nicer\ in lighter purple circles).}
    \label{fig:x-ray}
\end{figure*}


\label{sec:obs}

\subsection{HST COS and STIS}

The HST observing program (GO Proposal ID 16196) began a 1-year campaign on 2020-11-24. Using the Cosmic Origins Spectrograph (COS; \citealt{Green12}), we obtain medium resolution UV spectra covering 1070--1750 \AA\ in one-orbit visits scheduled on an approximately two-day cadence. Each visit consists of four 60~s exposures with grating G130M at the 1222\AA\ 
central wavelength setting and all four focal-plane (FP-POS) positions, exposures of 175~s and 180~s using G160M at the 1533\AA\ central wavelength setting at two FP-POS locations, and two 195 s exposures with G160M at the 1577\AA\ central wavelength setting also at two FP-POS locations. The multiple grating settings and FP-POS locations permit us to sample the detector at four locations for each wavelength to correct for
flat-field anomalies, grid-wire shadows, and gain sag in the detector \citep{COS_IHB20}.
For our analysis we use the calibrated spectra from the
Mikulski Archive for Space Telescopes (MAST) as processed by version 3.3.10 of the COS calibration pipeline.
The pipeline assembles the individual exposures into three combined spectra for each grating setting. We then combine these three spectra using the IRAF task {\tt splice} to form the merged 1070--1750 \AA\ final product.

Our observing program also includes monthly observations of the flux standard WD0308$-$565 using the same grating settings, which will permit us to track the time-dependent sensitivity of COS more precisely. We are not yet incorporating these potential corrections into the data presented here, which have calibration accuracies of $\sim$5\% in absolute flux, $\sim$2\% in day-to-day reproducibility, and a wavelength accuracy of $\sim 5~\rm km~s^{-1}$\citep{COS_IHB20}.

Our first spectrum revealed unexpected, strong, broad, blueshifted absorption
lines similar to the obscuring outflows seen in other Seyfert galaxies such as Mrk 335 \citep{Longinotti13}, NGC 5548 \citep{Kaastra14}, NGC 985 \citep{Ebrero16} and NGC 3783 \citep{Mehdipour17}. To obtain a full range of spectral diagnostics for understanding the new
obscuring outflow in Mrk 817, we arranged our observing program to include additional COS exposures covering shorter wavelengths down to 940 \AA, together with the scheduled Space Telescope Imaging Spectrograph (STIS;  \citealt{Woodgate98}) spectra
covering longer wavelengths up to 1\,$\rm \mu m$.

The additional COS spectra were obtained during Visit 3N on 2020-12-18, simultaneously with the XMM-Newton observation. They comprised four 522 s exposures at all four FP-POS settings with grating G130M at the 1096 central wavelength setting. When combined with the standard G130M and G160M exposures, the merged spectrum
covers 940--1750 \AA, as shown in Fig.\ref{fig_cosfull}.
The STIS exposures all used the 52x0.5 arcsec slit.
With grating G230L and the NUV-MAMA, a 320 s exposure covered 1600--3150 \AA. For the longer wavelengths we used the STIS CCD and a series of of CR-SPLIT exposures at three points along the slit to eliminate hot and cold pixels from
the final spectra. Using grating G430L three 30 s exposures covered 2950--5700 \AA,
and three 30 s exposures with grating G750L covered 5300--10,200 \AA.

With a continuum flux at 1398 \AA\ of
$7.7 \times 10^{-14}$ erg cm$^{-2}$ s$^{-1}$ \AA$^{-1}$, Mrk 817 has so far in our campaign been substantially brighter than the mean historical flux of
$4.3 \times 10^{-14}$ erg cm$^{-2}$ s$^{-1}$ \AA$^{-1}$ at 1397 \AA\ \citep{Dunn06}.
Our spectrum shows a rich range of broad absorption lines--
\ion{S}{6} $\lambda\lambda$933,944,
\ion{C}{3} $\lambda$977, \ion{N}{3} $\lambda\lambda$989,991,
\ion{O}{6} $\lambda\lambda$1031,1037, \ion{P}{5} $\lambda\lambda$1117,1128,
\ion{C}{3}* $\lambda$1176,
\ion{N}{5} $\lambda\lambda$1238,1242, \ion{Si}{4} $\lambda\lambda$1393,1402,
\ion{C}{4} $\lambda\lambda$1548,1550,
and Lyman lines from $\alpha$ through $\delta$.

In addition to these new, broad absorption features, the other
intrinsic absorption lines in Mrk 817 show continuing variability.
The first observation with COS in 2009 by \cite{Winter11} showed changes in
several lines compared to prior HST observations.
In our observations, the absorption lines described by \cite{Winter11}
have weakened, and a new strong
system of features at $v_\mathrm{out} = -3720 \rm~km~s^{-1}$ has appeared.
We measured velocities relative to the preferred systemic redshift in the NASA Extragalactic Database\footnote{\url{ned.ipac.caltech.edu}}  (NED)
of z=0.031455 from \cite{Strauss88}.
This new system shows absorption exclusively by high-ionization ions:
\ion{S}{6} $\lambda\lambda$933,944,
\ion{O}{6} $\lambda\lambda$1031,1037, \ion{N}{5} $\lambda\lambda$1238,1242,
\ion{C}{4} $\lambda\lambda$1548,1550, Ly$\alpha$, and Ly$\beta$.

\subsection{XMM-Newton}

Upon discovery of the unexpected features in the COS spectra and reduced soft X-ray flux indicated by Swift and NICER, we requested a 135~ks XMM-Newton Director's Discretionary Time (DDT) observation (PI: N. Schartel). The observation (OBSID: 0872390901) took place on 2020-12-18. We reduced the data using the {\em XMM-Newton} Science Analysis System (SAS v. 19.0.0) and the newest calibration files. We started with the observation data files and followed standard procedures. The observations were taken in Large Window Mode. The source extraction regions are  circular regions of radius 35 arcsec centered on source position. The background regions are also circular regions of radius 35 arcsec, avoiding the detector edges where the instrumental copper line is most prominent. The observations were relatively clean of soft proton background flares, save for a few kiloseconds removed from the end of the observation, for a final exposure of 115~ks. The response matrices were produced using rmfgen and arfgen in SAS. The PN spectra were binned to a minimum of 25 counts per bin to enable the use of $\chi^2$ statistic. The 2020-12-18 PN observation is shown in purple in Fig.~\ref{fig:x-ray}, compared to the only archival XMM-Newton observation, a 15~ks observation taken in 2009 when the source was significantly brighter in X-rays (light grey; \citealt{Winter11}).  

The XMM-Newton/RGS data were reduced using the standard pipeline tool, {\sc rgsproc}, which produces the source, background spectral files, and instrument response files. We binned the RGS 1 and 2 spectra with `optimal binning' \cite{kaastra16}, and fit the combined spectrum with the Cash statistic. Upon inspection of the RGS spectra, we discovered that the spectrum was dispersed such that a nearby star falls on the RGS detector and contaminates the background direction. This causes artificial `absorption' features in the background-subtracted spectrum, and also means that the RGS continuum is somewhat contaminated. We account for this by allowing the RGS to have a different continuum level from the PN spectrum. Future \xmm\ observations of Mrk~817 during the AGN~STORM~2 campaign will have a different position angle to avoid this star. 


\subsection{NuSTAR}

The 2020-12-18 XMM-Newton DDT observation was taken concurrently with NuSTAR observations (PI: J. Miller; \nustar\ GO Cycle~6). These NuSTAR results were recently published in \citet{miller21}, and corroborate spectral modelling shown here to the XMM-Newton observations. Here, we use the publicly available NuSTAR observations that took place on 2015-07-25 for 21~ks (obsid: 60160590002).  The NuSTAR Level 1 data products were processed
with the NuSTAR Data Analysis Software (NUSTARDAS
v2.0), and the cleaned Level 2 event files were produced
and calibrated with the standard filtering criteria using the NUPIPELINE task and CALDB version 20200813. The source and background regions were circular regions of radius 60 arcsec. The spectra were binned in order to oversample the instrumental resolution by a factor of three and to have a signal-to-noise of greater than $3\sigma$ in each bin.

The resulting NuSTAR spectrum is shown in Fig.~\ref{fig:x-ray}.
In Section~\ref{sec:xray_results}, we fit the XMM-Newton spectra together with the archival \nustar\ spectrum. While not simultaneous with our current campaign, it does not appear that the hard X-ray flux has changed significantly, consistent with the idea that most of the variability is due to line-of-sight obscuration. In particular, note that the 2015 NuSTAR and December 2020 XMM-Newton observations overlap in the 6--10 keV range.

\subsection{NICER}

NICER started monitoring Mrk 817 on 2020-11-28 (HJD~2459181) with an approximate cadence of every other day as part of a TOO request (PI: E. Cackett, Target ID: 320186). The data were processed using NICER data-analysis software version 2019-05-21\_V006 and CALDB version xti20200722 with the energy scale (gain) version ``optmv10''.
Short duration (typically $<100$~s) background flares were filtered out by excluding time intervals with a 13-15 keV count rate greater than 0.12~c\,s$^{-1}$, and spectra for each observation were constructed with the background estimator known as 3C50 (Remillard et. al, submitted).  Events were screened for overall high background rates using the level 3 filtering described in Remillard et. al, rejecting 1 observation. 14 observations with a mean overshoot rate higher than 0.28, corresponding to a high particle background which could not be modelled with confidence and were not identified by the 3C50 method, were also rejected.

The remaining 33 observations were divided into five time intervals, and the background-subtracted spectra within each interval were combined using addspec. Each combined spectrum was grouped using the ‘optimal binning’ scheme \citep{kaastra16}, with a minimum of 25 counts per bin. Two NICER spectra from towards the beginning and end of the time period we consider here are shown in Fig.~\ref{fig:x-ray}, demonstrating the evolution of the soft X-ray emission.

\subsection{Neil Gehrels Swift Observatory}
Swift \citep{gehrels04} began monitoring Mrk 817 on 2020-11-22 (HJD~2459175), with an approximate cadence of 1 day, aside from occasional gaps due to poor visibility from orbital pole constraints or interuptions caused by Gamma-ray Bursts or other Targets of Opportunity.  For this initial period of the campaign, data were taken as part of a TOO request (PI: E. Cackett, Target IDs: 37592 and 14012 were used). Each visit is typically $\sim$1 ksec. The Swift XRT \citep{roming05} was operated in Photon Counting mode. The UVOT \citep{burrows05} was typically operated in a end-weighted filter mode (0x224c) to get exposures in all 6 UV/optical filters with a weighting of 3:1:1:1:1:2 (for UVW2 through V), with occasional use of the blue-weighted 4-filter 0x30d5 mode (3 UV filters plus $U$) when shorter exposures are required close to periods of pole constraints.

Swift X-ray light curves were generated using the Swift-XRT \citep{evans07,evans09} data product tool\footnote{\url{https://www.swift.ac.uk/user_objects/index.php}}.
All archival and new UVOT data were processed and analyzed following the procedures described by \citet{edelson19} and \citet{Hernandez2020}, with HEASOFT version 6.28 and CALDB version 20210113.  Fluxes are measured using the uvotsource tool, with a circular source extraction region of 5\arcsec\ radius and with the background measured in a surrounding 40\arcsec--90\arcsec\ annulus. We apply detector masks to reject data points when the source falls on regions of the chip with lower sensitivity.  We follow the procedure laid out in \citet{Hernandez2020}, but find that the detector masks employed there are too aggressive for the present data, eliminating many points that are consistent with the light curves to within their measurement errors.  Instead, we use a more conservative set of masks defined by applying higher thresholds to the sensitivity maps, which results in eliminating 55 exposures from a total of 424. 

The Swift X-ray count rate during the campaign is significantly lower than seen previously \citep[e.g.,][]{morales19}.  The mean 0.3 -- 10 keV count rate between 2017 -- 2019 is 0.64~c\,s$^{-1}$, while during our campaign the mean rate is 0.077~c\,s$^{-1}$, a factor of 8 lower.  Despite the significant change in the X-ray count rate, the mean $UVM2$ flux is approximately same from 2017 -- 2019 as it is during our current campaign ($3.88\times10^{-14}$ vs. $3.96\times10^{-14}$ erg~s$^{-1}$~cm$^{-2}$~\AA$^{-1}$).  The 2017-2019 data show the X-ray and $UVM2$ rates are not correlated \citep{morales19}, and this continues in the current campaign.

\begin{deluxetable*}{llccl}
\label{table:optimg}
\tablecaption{Ground-based Imaging Observations}
\tablehead{
\colhead{Observatory/Telescope} & \colhead{Instrument} & \colhead{Filters} & \colhead{$N_\mathrm{epochs}$} & \colhead{References} }
\startdata
Calar Alto 2.2m & CAFOS & $V$ & 34 & \\
LCO 1m       & Sinistro & $BV$, $u^\prime g^\prime r^\prime i^\prime z_\mathrm{s}$ & 80 & \citet{Brown2013} \\
Liverpool Telescope 2m & IO:O & $u^\prime g^\prime r^\prime i^\prime z^\prime$ & 5 & \citet{Steele2004} \\
Wise Observatory 18-inch & QSI683 &  $g^\prime r^\prime i^\prime z^\prime$ & 66 & \citet{Brosch2008} \\
Yunnan Observatory 2.4m & YFOSC & $V$ & 28 & \\
Zowada Observatory 20-inch & & $g^\prime r^\prime i^\prime z_\mathrm{s}$ & 80 & \\
\enddata
\end{deluxetable*}

\subsection{Ground-based photometry}

Ground-based optical imaging was obtained using the facilities listed in Table \ref{table:optimg}. These include the 2.2 m telescope at Calar Alto Observatory in Spain, the 1 m telescopes of Las Cumbres Observatory Global Telescope Network (LCO) located at McDonald Observatory in Texas, the 2 m Liverpool Telescope located on the island of La Palma in the Canary Islands, the Wise Observatory Centurion 18-inch telescope (C18) in Israel, the Yunnan Observatory 2.4 m telescope in China, and the 20-inch telescope of Dan Zowada Memorial Observatory in New Mexico. Filters included Johnson or Bessell $B$ and $V$, SDSS $u^\prime g^\prime r^\prime i^\prime z^\prime$, and Pan-STARRS $z_\mathrm{s}$. Exposure times ranged from 10 s to 300 s, and at some facilities, two exposures per filter were taken on each observing night.

Basic processing steps including bias subtraction and flat-fielding were performed using the standard pipelines for each facility. For all of the data other than imaging from Calar Alto and Yunnan Observatory, photometry was carried out using an automated procedure written as a wrapper to routines in the photutils package \citep{larry_bradley_2020_4044744} of Astropy \citep{Astropy2018}. The procedure automatically identifies the AGN and a set of comparison stars in each image based on the object coordinates, locates the object centroids, and then performs aperture photometry using a 5\arcsec\ aperture radius and a sky background annulus spanning 15--20\arcsec. Scale factors are applied to the count values in order to minimize the scatter in the comparison star light curves (separately for each telescope), and comparison star magnitudes from the APASS catalog \citep{APASS_DR10} are used to calibrate the flux scale for each filter. For a given filter, data points from the same telescope and same night are combined using a weighted average. Finally, the separate light curves from each telescope are intercalibrated using PyCALI\footnote{\url{https://github.com/LiyrAstroph/PyCALI}.} \citep{Li2014}, in order to account for differences in wavelength-dependent throughput.  Specifically, PyCALI models light curves using a damped random walk process and applies additive and multiplicative factors to each telescope's data so as to align the fluxes into a common scale. To account for any systematic errors beyond the statistical uncertainties from the aperture photometry measurements, PyCALI also expands the error bars by adding a systematic error term in quadrature to the original uncertainties of each telescope's data. These intercalibration factors are determined in a Bayesian framework with a diffusive nested sampling algorithm (\citealt{Brewer2011}).

Photometry on the $V$-band data from Calar Alto and Yunnan Observatory was carried out with a separate software pipeline, using an aperture of radius 2\farcs7 and a sky background annulus of 5\farcs3--8\farcs0 for Calar Alto and an aperture radius of 5\farcs7 and a background annulus spanning 11\farcs4--17\farcs0 for Yunnan. These data points were merged with data from other telescopes using PyCALI to produce the final $V$-band light curve.

We combined the Pan-STARRS $z_s$ and SDSS $z^\prime$ data together into the final $z$-band light curve. Given the relatively lower S/N of the $z$-band data (partly due to CCD fringing noise), we do not find significant differences in light curve shape or reverberation lag between these two $z$-band filters.
In the following discussion, we will refer to the SDSS and Pan-STARRS filters as the \emph{ugriz} bands.

\begin{deluxetable*}{llcccc}
\label{table:optspec}
\tablewidth{0pt}
\tablecaption{Ground-Based Spectroscopic Observations}
\tablehead{
\multicolumn{1}{l}{Telescope} &
\multicolumn{1}{l}{Instrument} &
\colhead{Number of} &
\colhead{Wavelength} &
\colhead{Wavelength} &
\colhead{Aperture} \\
&
&
\colhead{Epochs} &
\colhead{Dispersion} &
\colhead{Coverage} &
\colhead{(Slit width $\times$}  \\
&
&
&
\colhead{(\AA\,pixel$^{-1}$)} &
\colhead{(\AA)} &
\colhead{extraction window)}
\\
}
\startdata
Calar Alto 2.2 m & CAFOS & 34 & 
4.47 &
4000--8500\phn & 
$3\arcsecpoint0 \times 10\arcsecpoint6$ \\
LCO 2 m & FLOYDS \citep{Brown2013} & 47 & 
3.51 &
5400--10000 & 
$2.0\arcsec \times 8\arcsec.8$\\
& & & 1.7 &
3200--5700\phn & 
$2.0\arcsec \times 8\arcsec.8$\\
Lick 3 m & Kast Spectrograph & 7 & 
1.0 &
3620--5700\phn & 
$4\arcsecpoint0 \times 15\arcsec$\\
 & & & 2.6 & 5700--10700 & 
 $4\arcsecpoint0 \times 15\arcsec$\\
Yunnan 2.4 m  & YFOSC & 8 & 
1.8 &
3800--7200\phn & 
$2\arcsecpoint5 \times 8\arcsecpoint5$\\
Liverpool& SPRAT \citep{Piascik2014} & 18 & 
4.6 &
4000--8000\phn & 
$1\arcsecpoint8 \times 3\arcsecpoint5$ \\
WIRO 2.3 m & Long Slit Spectrograph & 62 & 
1.49&
4000--7000\phn & 
$5\arcsec \times 13\arcsecpoint6$\\
ARC 3.5~m & TripleSpec & 28 & 
1.9 & 9500--24600 & 
$1\arcsecpoint1 \times 12\arcsecpoint0$ \\
Gemini North 8~m & GNIRS & 17 & 
2.5 & 8500--25000 & 
$0\arcsecpoint45 \times 1\arcsecpoint6$\\
IRTF 3~m & SpeX & 5 & 
2.5 & 7000--25500 & 
$0\arcsecpoint3 \times 0\arcsecpoint9$\\
\enddata
\end{deluxetable*}

\subsection{Ground-Based Optical
Spectroscopy}
\label{section:opticalspectroscopy}
The coordinated program of ground-based optical spectroscopy includes observations at six observatories: the 2.2 m telescope at Calar Alto Observatory, the 2 m Faulkes Telescope North of the LCO network on Maui, the 3 m Shane Telescope at Lick Observatory, the 2.4 m telescope at Yunnan Observatory, the 2 m Liverpool Telescope (LT), and the 2.3 m Wyoming Infrared Observatory (WIRO).
Details of the instrumental setups including wavelength coverage, dispersion, slit width, and extraction aperture are listed in Table \ref{table:optspec}.

Each set of data was processed independently using standard procedures for bias subtraction, flat-field correction, and cosmic ray removal. Spectroscopic extractions and calibrations were carried out separately for each instrument, applying methods as described for each facility: Calar Alto \citep{Hu2020}; LCO (the AGN FLOYDS pipeline\footnote{\url{https://github.com/svalenti/FLOYDS_pipeline}}); Lick \citep{Silverman2012}; Yunnan \citep{Du2014}; LT (the SPRAT data reduction pipeline\footnote{\url{https://telescope.livjm.ac.uk/TelInst/Inst/SPRAT}}); WIRO \citep{Brotherton2020}.

Flux calibration of the spectra from Calar Alto and Yunnan was done using
a comparison star observed simultaneously in the rotated slit as described by 
\citet{Hu2021}. For data from the other telescopes, the spectra were scaled
to have a constant [\ion{O}{3}] $\lambda$5007 emission-line intensity with the
spectral fitting method described in \citet{Hu2016}. For each telescope, the \hb\
flux was measured by integration over the range 4951--5075 \AA\ in
the observed frame, relative to a continuum defined in the windows 4879--4930
\AA\ and 5235--5286 \AA\ (the 5100 \AA\ continuum). The 5100 \AA\
continuum and \hb\ light curves from different telescopes were intercalibrated
as per Equations (1) and (2) in \citet{Peterson2002} to correct for aperture
effects, and then measurements with observation times closer than 0.8 days were averaged.

\subsection{Ground-Based Near-IR Spectroscopy}
\label{section:nearIRspectroscopy}

In AGN STORM 1, NGC~5548 was monitored with near-IR spectroscopy two years after the main campaign \citep{landt2019}. For AGN STORM 2 we made a point of obtaining contemporaneous NIR spectroscopy using Gemini North 8~m, ARC 3.5~m and IRTF 3~m (see Table~\ref{table:optspec}). Details of these observations will be presented in a follow-up paper, but we note here that there is no evidence for absorption in the low-ionization line \hei~$1.08$~$\mu$m (as of 2021 April 3) from either the narrow or the broad UV absorbers. This situation is very different from that of NGC~5548, which showed \hei~$1.08$~$\mu$m absorption from both the `obscurer' and warm absorber \citep{wildy2021}.

\section{Results}

\subsection{Spectral modelling}
\label{sec:dec2020}

We begin with a close examination of our highest quality observations taken on 2020-12-18 with \HST\ and XMM-Newton. In the following section, we describe the spectral fits, which will be the baseline models for the time-resolved spectroscopy in Section~\ref{sec:evolution}, and the broadband SED and ionic column densities used for photoionization modelling in Section~\ref{sec:photoionization}.

\subsubsection{The UV spectrum from 2020-12-18}\label{sec:uv_results}

The rich set of UV absorption lines permits a detailed investigation of the
properties of the absorbing gas.
The many doublets and the Lyman series allow us to obtain good measurements of
the ionic column densities and covering fractions due to their range in oscillator strengths.

To measure the absorption features in the HST spectra,
we first model the emission spectrum.
Our approach is similar to that used by \cite{Kriss19b} in their analysis of NGC~5548
spectra from AGN STORM 1.
Our model is empirical, and although the components of the model
are not strict physical representations, they enable us to
separate the emission lines, the absorption features, and the continuum.
Although we use an accretion disk spectrum to model the broad-band continuum
of Mrk 817, to fit the far-UV spectra in detail, we approximate the
940--1750 \AA\ continuum with a reddened powerlaw,
$F_\lambda(\lambda)  = F_\lambda(1000\, \mbox{\AA}) (\lambda / 1000\,\mbox{\AA})^{- \alpha}$,
assuming foreground Galactic extinction with a color excess $E(B - V) = 0.022$ \citep{Winter11},
and a ratio of selective to total extinction of $R_V = 3.1$.
Although NED suggests $E(B - V) = 0.005$, we use the higher value from
\cite{Winter11} because it provides a better spectral fit.
There is no indication of additional extinction in Mrk 817.
All spectral components are absorbed by foreground damped Ly$\alpha$ by
Galactic neutral hydrogen with a column density of
$N$(\ion{H}{1})$ = 1.15 \times 10^{20}~cm^{-2}$ \citep{Murphy96}.

For the emission lines, we use multiple Gaussian components.
The brighter lines (Ly$\alpha$, \ion{C}{4}, \ion{O}{6}) require four
components associated with each individual multiplet, ranging from a weak
narrow-line component of width $\sim500~\rm km~s^{-1}$ to a very broad
component with width of $\sim12000~\rm km~s^{-1}$.
Weaker lines such as \ion{P}{5}, \ion{Si}{2}, and \ion{C}{2} require only a
single broad component.


\begin{figure*}
    \centering
    \begin{minipage}{.49\textwidth}
        \centering
        \includegraphics[width=.95\linewidth]{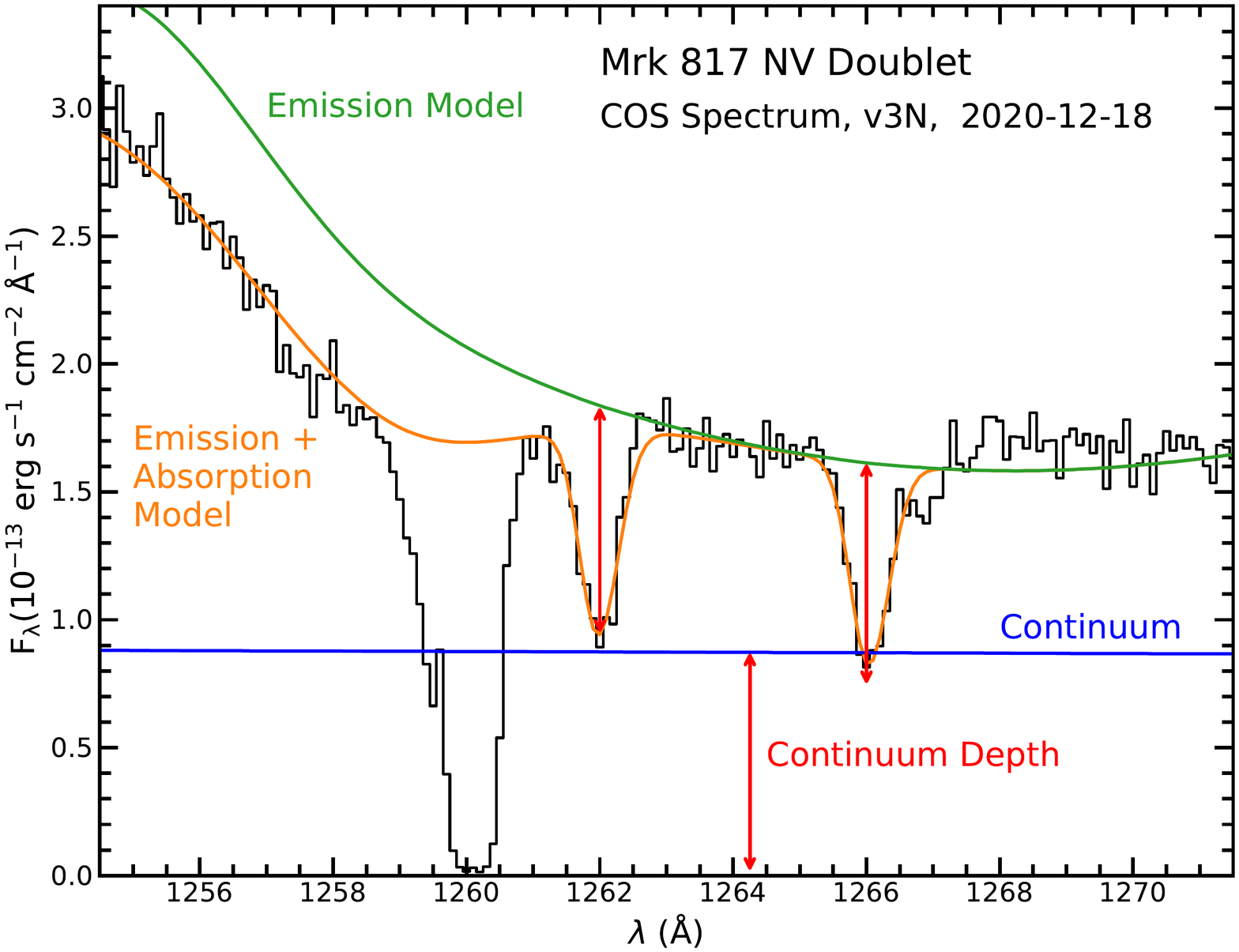}
    \end{minipage}%
    \begin{minipage}{.49\textwidth}
        \centering
        \includegraphics[width=.95\linewidth]{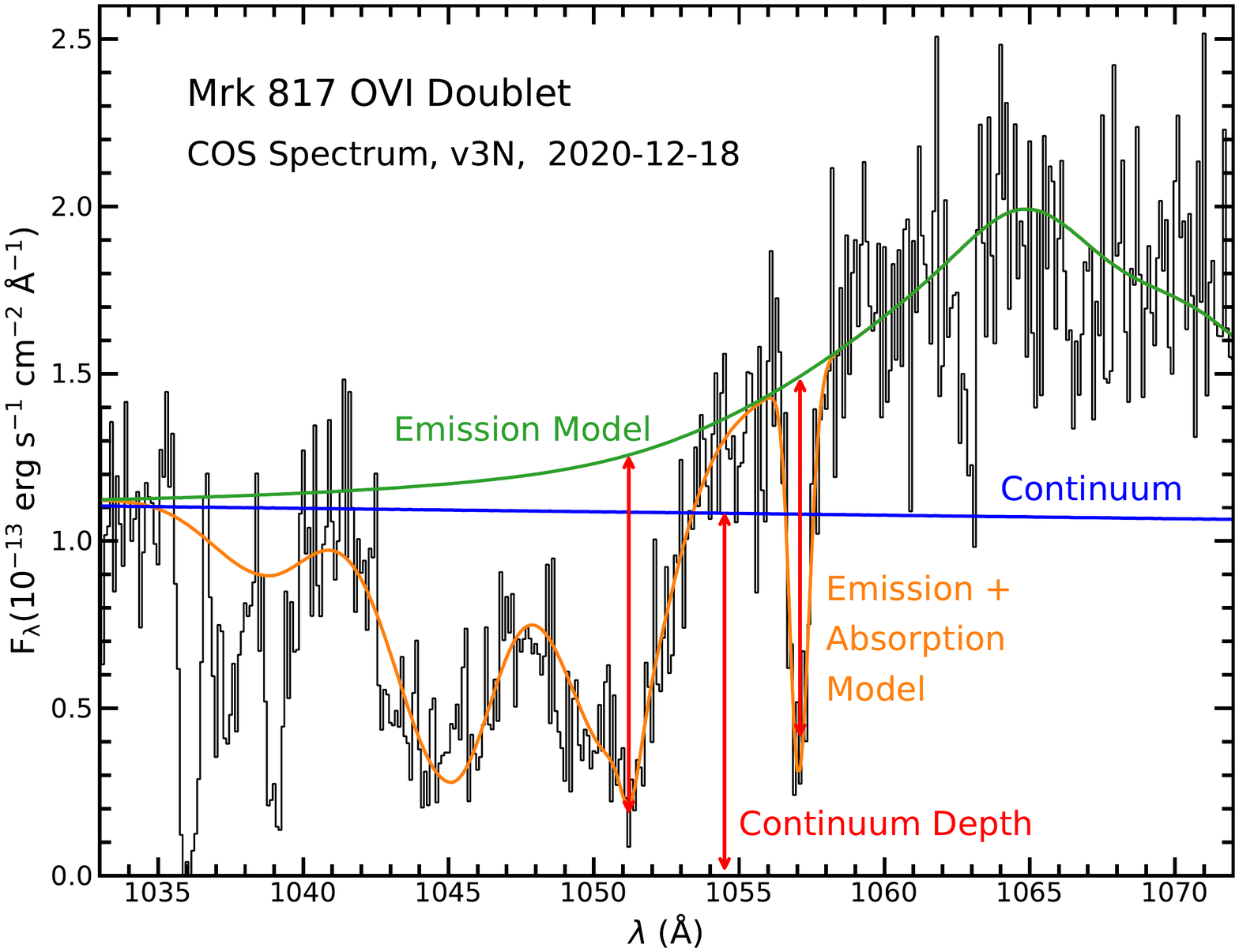}
    \end{minipage}%
\caption{{\em Left:} COS spectrum of the region surrounding the \ion{N}{5} absorption
doublet in Mrk 817 from 2020-12-18 (black).
The data are binned by 8 pixels ($\sim$1 resolution element) and are plotted as a function of
observed wavelength.
The best-fit emission model is in green, and the best-fit model including both
broad and narrow absorption is in orange.
The model continuum is in blue.
The red arrow represents the continuum depth, and we compare it to the depths of
both the blue and the red transitions as measured from the total emission model.
Note that the blue line has the same depth as the continuum, but that the
red line is slightly shallower. The deep trough bluewards of \ion{N}{5} is due to foreground interstellar lines: \ion{Si}{2} 1260.42\AA\ (strongest line) and \ion{S}{2} 1259.52\AA\ (weak dip on the blue side). {\em Right:} Same as left, but for the region surrounding the broad and narrow
\ion{O}{6} absorption doublets. Both the blue and the red transitions have the same depth as the continuum, 
indicating saturation and full coverage of the continuum emission.
}\label{fig:uv_abs}
\end{figure*}


We use negative flux Gaussian profiles to model the intrinsic absorption
components.
Single narrow Gaussians with a full-width at half-maximum of
FWHM$\sim150~\rm km~s^{-1}$ model the narrow absorption lines well.
For the broad absorption troughs, we permit these Gaussians to be
asymmetric with different dispersions on the blue and red sides of the profile, i.e. for a line centered at $\lambda_c$, the Gaussian dispersion for $\lambda > \lambda_c$ is $\sigma_\mathrm{red}$, and for $\lambda < \lambda_c$ it is $\sigma_\mathrm{blue}$.
The ratio of the dispersions is a free parameter. 
The broad troughs in Mrk 817 tend to have a rounded triangular profile with the
deepest point of the trough on the blue side and a more extensive wing on the
red side. Their widths range from $\sim1000-1500~\rm km~s^{-1}$.
Note that this is not a physical description for the absorbing gas.
Deriving physical information (e.g. column densities and covering fractions)
requires making further assumptions on the origin of the velocity profile
of an absorption trough and on which elements of the emission model are actually
covered by the absorbing gas.

The narrow absorption lines are simple, resolved, and symmetric.
Aside from \ion{O}{6} they appear to be unsaturated.
The broad absorption troughs are more complex.
They appear to be fully saturated, with profiles determined by variations in
the covering factor as a function of velocity.
Deep troughs like Ly$\alpha$, \ion{C}{4}, \ion{O}{6}, and \ion{S}{6} have
almost exactly the same asymmetric profile shape-- steeper on the
blue side, shallower on the red side.
Even lines of low abundance elements like \ion{P}{5} have doublet ratios of $\sim$1:1, and so are saturated, even when the line has a lot equivalent width.
We point readers to Appendix Table~\ref{tab_abs_lines} for a summary of the equivalent width (EW) and other empirical properties of the narrow and broad absorption lines from observations on 2020-12-18.

To determine whether the absorbing gas in each case
covers all the emission (e.g.\ lines and continuum), or just the continuum, or
fractions of each, we look in detail at the narrow \ion{N}{5} and \ion{O}{6}
doublets. Fig.~\ref{fig:uv_abs}-{\em left} shows that the red
transition of the \ion{N}{5} doublet is deeper than the blue
transition.
However, our full emission model (that accounts for the broad \ion{N}{5}
absorption trough) shows that, compared to the unabsorbed overlying
emission, the blue transition is indeed slightly deeper.
In fact, it has exactly the same depth as the continuum.
Although this is not a definitive interpretation, it implies
that the narrow emission covers only the continuum and not the broad-line
emission.

Similarly, Fig.~\ref{fig:uv_abs}-{\em right} illustrates the same behavior for \ion{O}{6}.
In this case the blue transition of the \ion{O}{6} doublet falls within the
red trough of the broad \ion{O}{6} absorption.
At first glance, the red transition of the doublet again appears deeper
than the blue, but when the depth of the blue line is measured relative to
the full emission model, we see that both lines are consistent with
saturation at full coverage of only the continuum.
The residual emission below the bottom of the blue line equals the modeled
intensity of the overlying broad-line emission as if that emission
is unabsorbed.

Similar arguments apply to the broad \ion{O}{6} absorption troughs.
Relative to the continuum emission, they both have the same depth.
However, that depth is shallower than the continuum strength, indicating that
the broad lines are saturated, but that they only partially cover the continuum.

We do not show the same level of detail for \ion{C}{4}, \ion{Si}{4},
\ion{P}{5}, or \ion{S}{6}, but Fig.~\ref{fig_cosfull} shows that the regions
surrounding these doublets are strongly dominated by continuum emission.
We therefore measure column densities assuming that both the broad and narrow
absorption lines cover only the continuum emission.

For \ion{S}{6} the depths of the blue and red transitions have close to a 2:1
ratio and appear to be nearly optically thin.
The broad absorption troughs in \ion{S}{6} have equal depths and appear to be
nearly saturated, but they only partially cover the continuum.
Like \ion{S}{6}, the broad troughs in \ion{C}{4} and \ion{P}{5} appear to be
saturated, and only partially cover the continuum.
\ion{Si}{4} shows only broad absorption troughs.
The red transition in \ion{Si}{4} is shallower than the blue, indicating that
the doublet is unsaturated. This yields a reliable measure of both the column
density and the covering factor using the doublet method of \cite{Barlow97}.

Table~\ref{tab_columns} gives our measurements of the ionic column densities
for both the narrow and the broad absorption troughs, using the apparent optical depth (AOD) method of \citet{Savage91}.
We assume they both only cover the continuum and not the overlying broad
line emission.
For the narrow absorption lines, we assume that the
covering factor is 100\%, $f_c=1$, and uniform in velocity.
For each line we give a best measured value, and quote a $2\sigma$ lower limit ($\Delta \chi^2 = 3.82$) from the best-fit $\chi^2$ value
or the value from direct integration of the line profile normalized by the
continuum emission, whichever is lower.
Upper limits are also reported at $2\sigma$ ($\Delta \chi^2 = 3.82$).

Just as the \ion{Si}{4} doublet appears to be unsaturated, the highest order
Lyman lines are also unsaturated, giving us a well constrained series solution
for the \ion{H}{1} column density.
Ly$\alpha$ is highly saturated, with its profile determined by covering factor.
Assuming the same profile shape for the higher-order lines, we fit the series
in a consistent way.
Measuring column densities independently for each line using these fitted
profiles and assuming the same covering factor as a function of velocity,
$\rm f_c(v)$, as derived for Ly$\alpha$ gives consistent \ion{H}{1} column
densities as shown in Table \ref{tab_h1}.

For singlets like \ion{C}{3}* $\lambda$1176, \ion{C}{3} $\lambda$977 and
\ion{N}{3} $\lambda$991, we again assume their broad absorption troughs
are saturated.  We use the shape of Ly$\alpha$ to determine their profiles
and assume that the trough shape determines the covering factor as a function
of velocity, $f_c(v)$.
To measure the column density, $N_\mathrm{ion}$,
we integrate the line profile using $f_c(v)$ from Ly$\alpha$ and assume an optical depth of $\tau=4.5$ at its deepest point, which corresponds to a
$\sim 1$\% residual intensity. (Given the high degree of saturation, these column
densities could be even higher.)

For blended multiplets like \ion{C}{3}* $\lambda$1176, \ion{N}{3} $\lambda$991,
and \ion{C}{4} $\lambda$ $\lambda$1548,1550, we sum the oscillator strengths
to convert the optical depth to a column density.
Given the high degree of saturation in many lines, column densities could be
even higher.  However, the unsaturated \ion{Si}{4} doublet and the
unsaturated higher order Lyman lines, Ly$\gamma$ and Ly$\delta$,
give secure column density measurements.

\subsubsection{The X-ray spectrum from 2020-12-18}\label{sec:xray_results}

\begin{figure}
\begin{center}
\includegraphics[width=\columnwidth]{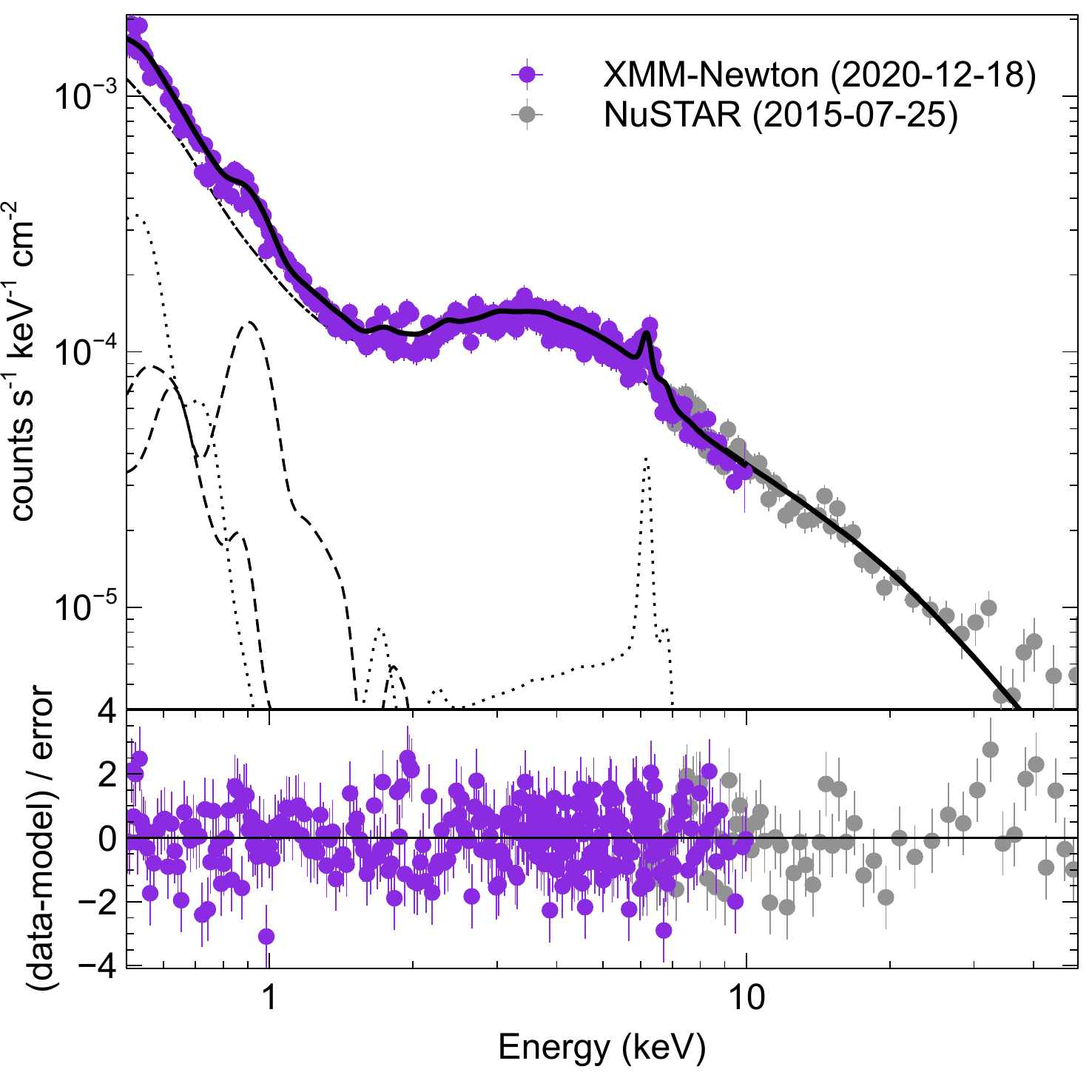}
\caption{The \xmm\ spectrum from 2020-12-18 (purple) and archival \nustar\ spectrum, overlaid with the best fit model described in Table~\ref{tab:xmm_fits} (black solid line). The model consists of several components: emission from the inner accretion flow (a Comptonization continuum and relativistic reflection) that is absorbed by the new ionized obscurer (black dot-dash line); distant reflection fitting the narrow iron~K line (dotted line); and photoionized emission from two zones of circumnuclear gas (dashed lines).}
\label{fig:x-ray_spec}
\end{center}
\end{figure}

As indicated in Fig.~\ref{fig:x-ray}, at the beginning of the AGN~STORM~2 campaign, Mrk~817 was in a deep X-ray low flux state. Compared to the archival \xmm\ and \nustar\ observations, most of the changes are in soft X-rays. This fact, together with the sudden appearance of broad UV absorption lines (including from low-abundance elements like phosphorous), lead us to the conclusion that the X-ray variability is largely due to obscuration of the central compact source. 

There are sharp features in the X-ray spectrum below 1~keV that do not vary between the \xmm\ spectrum and lowest flux \nicer\ spectrum from 2021-01-20, despite significant diversion at higher energies. This suggests emission from distant circumnuclear gas that responds on long timescales. To probe further, we examine the high-resolution RGS data covering the 10--36\AA~(0.35--1.2~keV) range. We fit the continuum with a phenomenological powerlaw + blackbody model (though more complex/physical models do not change the results). There are clear indications of emission lines from \ion{O}{8} \lya, \ion{O}{7} triplet, \ion{N}{7}, \ion{C}{6} radiative recombination continuum (RRC), and \ion{C}{6}  at longer wavelengths. These can be well fit in {\sc spex} \citep{kaastra96,kaastra20} with two {\sc pion} photoionization models \citep{Mehdipour2016},  where the velocity broadening was fixed to $\sigma=400$\,\kms. Using the \oviii\ \lya\ line (and confirmed by comparison with \nvii\ and \ovii\ resonance line), we measure an outflow velocity of $v=-400\pm70$\,\kms. We fix the outflow velocity of the {\sc pion} models at this value, and fix the density of the slab to 1 cm$^{-3}$, as it does little to influence the line flux. The most important difference between these two components is their ionization parameters. Given the quality of our data, we cannot distinguish between the covering factor, $\Omega$, and the column density, as both a higher column density and larger covering factor lead to an increase in line strength. The best fit column densities, ionization parameters and covering factors for these two components can be found in Table~\ref{tab:xmm_fits}. The RGS data will be the subject of follow-up papers. For now we simply use the results to inform our fitting of the lower-resolution broadband spectra (Fig.~\ref{fig:x-ray_spec}).

\begin{deluxetable}{ccc}
\tablecaption{X-ray spectrum model parameters}\label{tab:xmm_fits}
\tablehead{
\colhead{Component} & \colhead{Parameter} & \colhead{Value} }
\startdata
tbabs & $N_\mathrm{H}$ ($10^{22}$cm$^{-2}$) & $0.01\tablenotemark{c}$ \\
\hline
zxipcf & $N_\mathrm{H}$ ($10^{22}$cm$^{-2}$) & $6.95^{+0.8}_{-0.7} $ \\
& $\log \xi$ (erg$\cdot$cm$\cdot$s$^{-1}$)& $0.55^{+0.3}_{-0.4}$ \\
& Cov. Frac.~$\Omega$ & $0.93^{+0.01}_{-0.008}$ \\
\hline
\relxillD\ & index & $>6.6$\\
& $a_*$ & $>0.97$\\
& $i$ (degrees) & $<40$ \\
& $\Gamma$ & $1.91^{+0.04}_{-0.09}$\\ 
& $\log \xi$ (erg$\cdot$cm$\cdot$s$^{-1}$)& $2.7^{+0.2}_{-0.3}$\\
& $A_{\rm Fe}$ & $6^{+3}_{-2}$\\
& $\log N_{\rm e}$ (cm$^{-3}$) & $18.7^{p}_{-0.6}$\tablenotemark{b} \\
& reflection fraction & $0.3^{+0.3}_{-0.2}$ \\
\hline
{\sc pion} &  $\log \xi$ (erg$\cdot$cm$\cdot$s$^{-1}$)& $2.7\pm0.3$\tablenotemark{a} \\
& $N_\mathrm{H}$ ($10^{21}$cm$^{-2}$) & 7.6\tablenotemark{a} \\
& Cov. Frac.~$\Omega$ & 0.02\tablenotemark{a} \\
\hline
{\sc pion} &  $\log \xi$ (erg$\cdot$cm$\cdot$s$^{-1}$)& $1.5\pm 0.2$
\tablenotemark{a} \\
& $N_\mathrm{H}$ ($10^{21}$cm$^{-2}$) & 50.6\tablenotemark{a} \\
& Cov. Frac.~$\Omega$ & 0.011\tablenotemark{a} \\
\hline
\xmm\ & $\chi^2$/d.o.f. & 1083/1109\\
\nustar\ & $\chi^2$/d.o.f. & 353/314\\
\enddata
\tablenotetext{a}{Parameter fit to RGS spectrum and fixed for PN analysis}
\tablenotetext{b}{Pegged to maximum value of $\log N_{\rm e} = 10^{19}$ cm$^{-3}$}
\tablenotetext{c}{Fixed parameter}
\end{deluxetable}

 We fit the CCD spectra in {\sc xspec} \citep{arnaud96}. For the soft photoionized emission lines (dashed black lines; Fig.~\ref{fig:x-ray_spec}), we used a pre-computed and simplified table version of {\sc pion}, called {\sc pion\_xs} \citep{parker19}. The grid assumes that the ionization continuum is a powerlaw spectrum with $\Gamma=2,$ typical of AGN and appropriate for this source. The PN data are most sensitive to the ionization parameters, and so we freeze them to the best fit values from the RGS analysis. There is also a clear detection of a narrow neutral iron~K$\alpha$ line, which we fit with the {\sc xillver} model \citep{garcia10}, with the log of the ionization parameter fixed at 0 (dotted black line; Fig.~\ref{fig:x-ray_spec}). We model the continuum as a cutoff powerlaw ($E_{\mathrm{cut}}=300$~keV) absorbed by a partial covering ionized absorber ({\sc zxipcf} in {\sc xspec}; \citealt{reeves08}). This resulted in an adequate fit ($\chi^{2}/{\mathrm{dof}}=1451/1413$). Similarly good fits and parameter estimates are found when using the SPEX model {\sc pion}, which properly accounts for the ionizing continuum. The addition of a soft excess component below 1~keV improves the fit at the $3\sigma$ level. In the literature, the soft excess has often been modeled as a `warm corona' (e.g., {\sc comptt} in {\sc xspec}; \citealt{petrucci13}), or as relativistically broadened reflection (e.g., {\sc relxillD}; \citealt{dauser2010}).  In favor of the reflection interpretation, \citet{garcia19} argue that the optically-thick warm corona would produce a spectrum substantially different from a blackbody with electron scattering due to atomic absorption, and such features are not observed in AGN. Our spectrum of Mrk~817 cannot statistically differentiate between these models ($\chi^{2}/{\mathrm{dof}}=1436/1409$ using the {\sc comptt} model and $\chi^{2}/{\mathrm{dof}}=1437/1409$ for the {\sc relxillD} model). We choose to proceed with the relativistic reflection model here, and in Sections~\ref{sec:evolution} and \ref{sec:photoionization}. Most importantly, we see that the choice of soft excess model does not affect the inferred properties of the ionized obscurer. The final model in {\sc xspec} syntax is:
{\sc tbabs*(pion\_xs+ pion\_xs + xillver + zxipcf*(relxillD))}. See Table~\ref{tab:xmm_fits} for details of the fit.

Finally, we tested a very different scenario where instead of an ionized obscurer, the low flux state was caused by an intrinsically low-flux corona, where the corona is extremely close to the black hole and light bending effects cause most of the photons to either fall into the black hole or irradiate the inner ($\sim 2~r_{\mathrm{g}}$) accretion disk. Such a model has successfully explained the spectra and reverberation time lags of highly variable narrow-line Seyfert 1~AGN (e.g., 1H\,0707-495; \citealt{Fabian2012} or IRAS~13224-3809; \citealt{Kara2013,Alston2020}). This model does provide a reasonable fit to the \xmm\ data ($\chi^{2}/{\mathrm{dof}}=1118/1109$), but overpredicts the \nustar\ flux, and the joint \xmm\ and \nustar\ fit is poor. Moreover, in such a scenario, the newly discovered broad absorption troughs in the UV would not be due to a new obscuration event, but instead due to changes in the ionizing continuum. This is challenging to reconcile with observations of low-abundance species like phosphorous, which point instead to a high column density gas. Because of this and the statistical evidence of the X-ray fit, we do not consider this model further.

\subsection{The evolution of the ionized obscurer}
\label{sec:evolution}

\begin{figure}
\begin{center}
\includegraphics[width=\columnwidth]{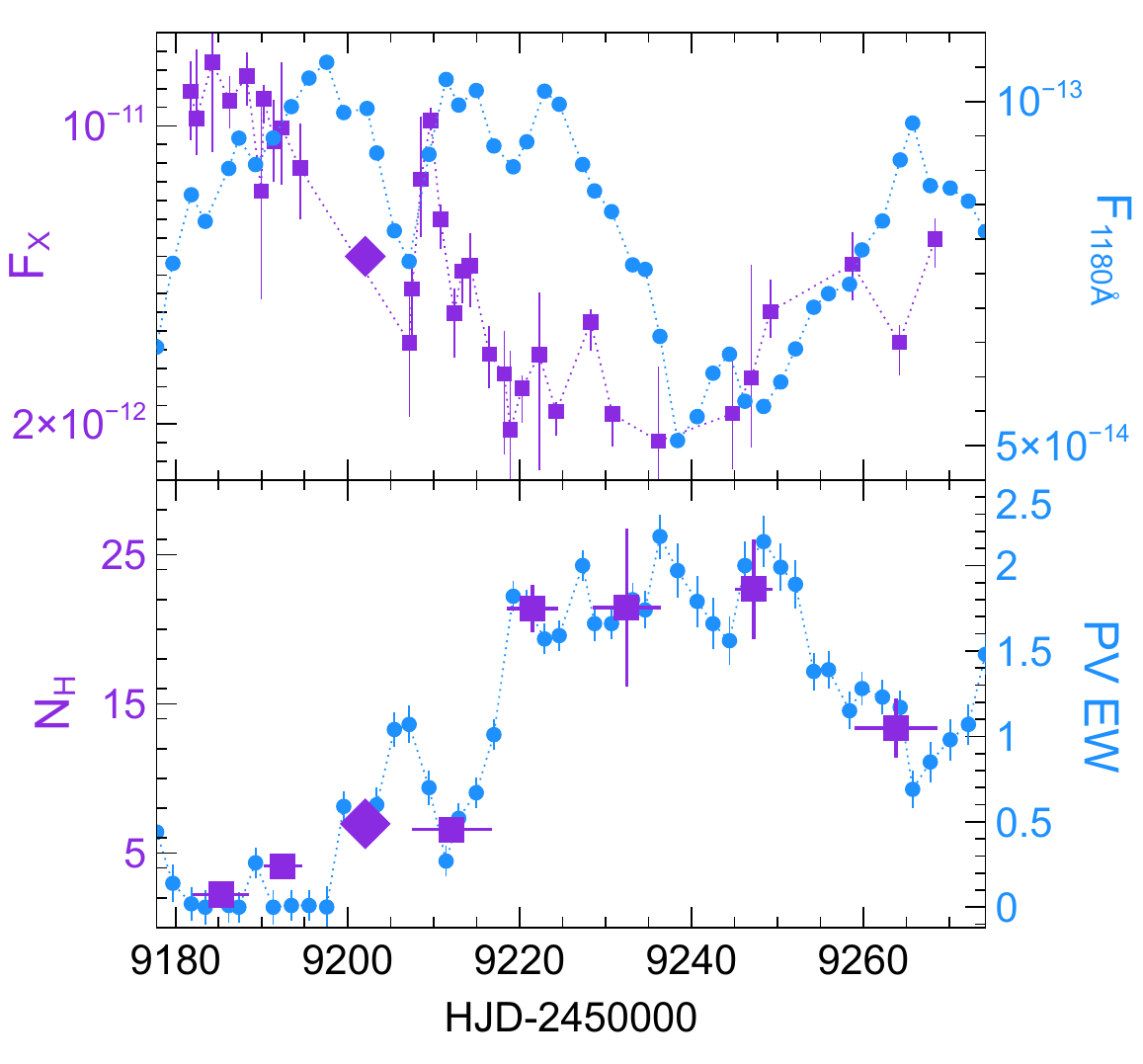}
\caption{Comparison of X-ray and UV absorption during the first third of the AGN~STORM~2 campaign. The top panel shows the \nicer\ 0.3-10~keV flux in units of erg~cm$^{-2}$~s$^{-1}$ (purple) compared to the 1180 \AA\ flux (erg~cm$^{-2}$~s$^{-1}$~\AA$^{-1}$) measured by HST (blue). The purple diamond demarcates the \xmm\ observation (also in 0.3-10 keV range). The bottom panel shows the column density of the X-ray obscurer in units of $10^{22}$ cm$^{-2}$ (purple), overlaid with equivalent width of the broad \ion{P}{5} absorption trough in units of \AA\ (blue). }
\label{fig:evolution}
\end{center}
\end{figure}

The \HST\ and \xmm\ observations presented above represent just one day of the year-long AGN~STORM~2 campaign.  Here we use the models presented above as the baseline for fitting the daily UV HST and NICER X-ray observations to track the evolution of the obscurer.

The \xmm\ spectral decomposition discussed in Section~\ref{sec:xray_results} provided constraints on the column density, ionization and covering factor of the ionized absorber that obscures the X-ray source. We use the best fit model in Section~\ref{sec:xray_results} as the baseline model, and freeze parameters that should not display inter-day variability (e.g., elemental abundances, black hole spin, emission from parsec-scale gas and beyond), and only allow the ionized absorber parameters (column density, ionization parameter, covering fraction) and continuum parameters (powerlaw photon index and normalization) to vary. To constrain these parameters, we needed to increase the signal to noise by binning the NICER data into 7 epochs, each now with $\sim 5-10$~ks of data. In the first 40 days of the campaign, the X-ray flux decreased by over an order of magnitude. Our time-resolved spectral analysis shows that this is largely driven by changes in the hydrogen column density of the ionized absorber (see Fig.~\ref{fig:evolution}).

We compare the X-ray absorber evolution to the broad P~V absorption line EW and ionizing continuum in Fig.~\ref{fig:evolution}. The same trend in EW and ionizing continuum is seen in all of the broad absorption troughs of lower ionization, lower abundance species, including \pv, \ciii* and \siIV. The bottom panel of Fig.~\ref{fig:evolution} shows a clear correlation between the column density measured in X-rays and the strength of the broad absorption trough.

\subsection{Photoionization Modeling}
\label{sec:photoionization}

To constrain the location and composition of the absorber producing the narrow and broad UV absorption lines shown in Section~\ref{sec:uv_results}, we used Cloudy photoionization models. The ionization and thermal equilibrium in a photoionized plasma is dependent on the spectral energy distribution (SED) of the ionizing source. To construct a model for the ionizing SED, we fit the HST COS+STIS spectrum and the accompanied XMM-Newton observation taken in 2020-12-18.

\subsubsection{The ionizing SED}

%
\begin{figure}[!tbp]
\centering
\resizebox{\hsize}{!}{\includegraphics[angle=0]{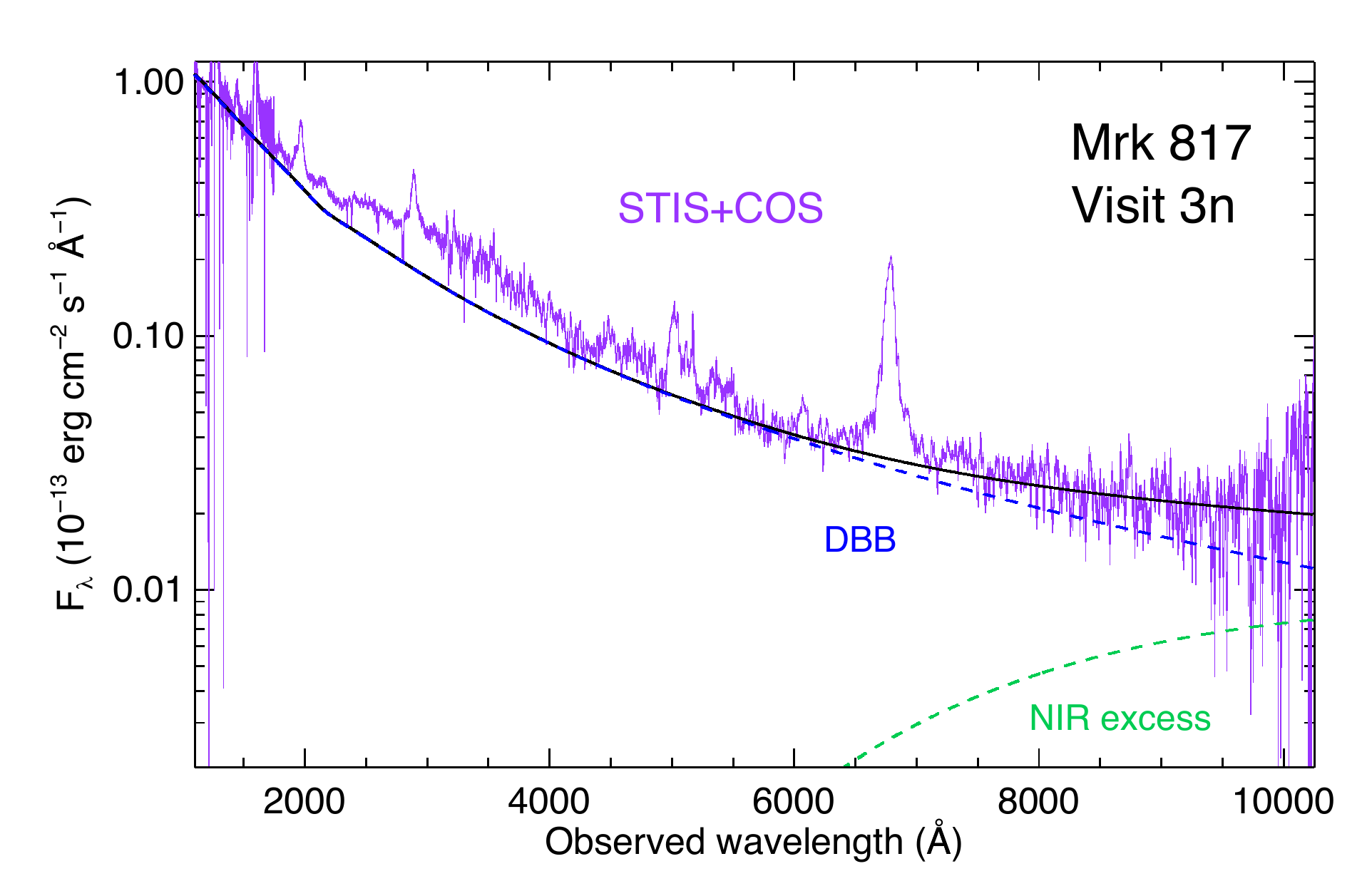}}
\caption{NIR-optical-UV spectrum, taken with HST COS and STIS on 2020-12-18. The continuum is fitted with a disk blackbody (DBB; dashed blue line). The NIR excess is modeled with a simple blackbody component (dashed green line) and may be attributed to thermal emission from the hottest regions of the dusty torus. The excess emission above the DBB continuum in the 2000--5500 \AA\ range is the complex \ion{Fe}{2} and Balmer continuum emission, which we exclude in our fitting of the continuum. The total best-fit continuum model is shown in solid black line, which is reddened with $E(B-V)=0.022$.}
\label{fig:SED_cos}
\end{figure}

%
\begin{figure}[!tbp]
\centering
\resizebox{\hsize}{!}{\includegraphics[angle=0]{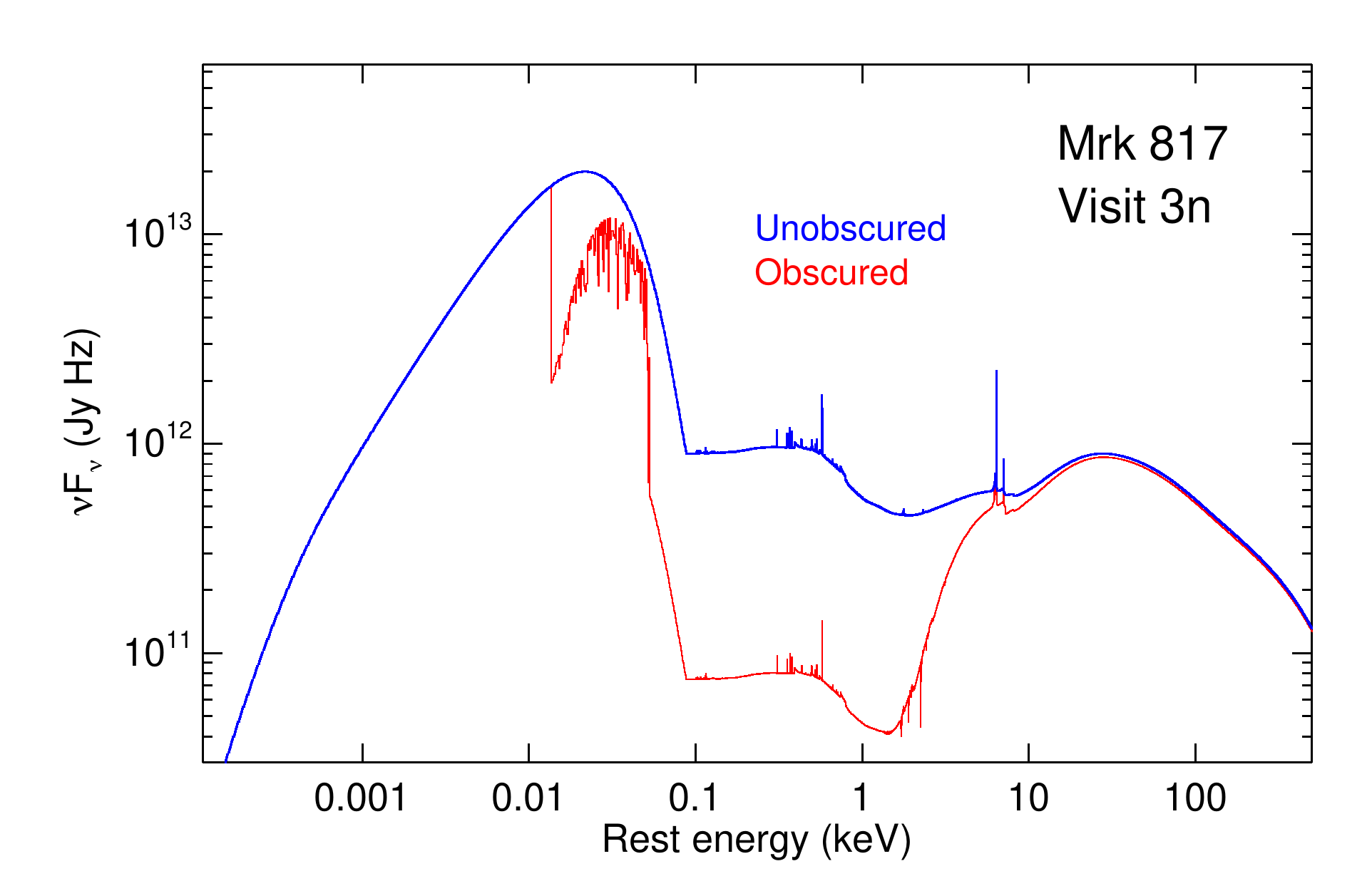}}
\caption{SED model of Mrk~817, before and after adding absorption by the obscuring gas. The unobscured SED (blue) has a luminosity of $1.1\times 10^{45}$~erg~s$^{-1}$ ($L/L_{\mathrm{Edd}}=22$\%), and the obscured (red) has $6.7\times 10^{44}$~erg~s$^{-1}$.}
\label{fig:SED_comparison}
\end{figure}

To model the NIR to UV continuum (Fig.~\ref{fig:SED_cos}), we used a disk blackbody component (DBB in SPEX), which provides a good fit to the continuum of the STIS+COS spectrum for a maximum temperature ${T_{\rm max} = 20}$~eV. This model is based on a geometrically thin, optically thick, Shakura-Sunyaev accretion disk. Reddening was accounted for in the same way as in Section~\ref{sec:uv_results}. We observe an excess of emission above the DBB continuum over the 2000--5500 \AA\ range, typical of the \ion{Fe}{2} and Balmer continuum emission of most AGN. This is notoriously challenging to model accurately, and so we exclude this emission feature in our fitting of the continuum. In order to obtain a good fit to the optical-UV continuum with the DBB model, we find that we are left with some excess emission in the NIR above the DBB continuum at $> 7000$ \AA. We attempted different models for this excess emission. We tested the various bulge and host galaxy starlight template models of \cite{Kin96}. However, such starlight models over-predict the 5500--6000 \AA\ continuum observed in the STIS spectrum. We find the NIR excess is best modeled as the short wavelength Wien tail from the hottest regions of the dusty torus. We thus used a simple blackbody component for this excess emission. The presence of such a torus contribution in the NIR is also supported by the 2MASS $J$, $H$, $K_s$ fluxes that are reported in NED (the `profile-fit' values of \citealt{Skru06}). While a Wien tail is our favored description, we cannot rule out other origins of this blackbody component (e.g. diffuse continuum emission \citealt{chelouche19,netzer20}). Regardless, it does not affect our photoionization modeling.

The model for the X-ray to EUV spectrum is based on results presented in Section~\ref{sec:xray_results}, where the soft excess is fit by relativistic reflection off the inner accretion disk (see Section~\ref{sec:xray_results} for details). We also consider photoionization models that assume a warm corona soft excess (similar to the SED assumed in \citealt{Mehdipour2015}), and find essentially the same solution. The ionized absorber is modeled with PION in SPEX because it extends down to the Lyman limit. The parameter constraints are similar to those found with {\sc zxipcf} in earlier sections: $N_\mathrm{H} = 9.6\pm0.6 \times 10^{22}$~cm$^-2$, $\log \xi = 1.0 \pm 0.6$~(erg~cm~ s$^-1$) and covering fraction $\Omega = 0.92 \pm 0.01$. The 1--1000~Ryd ionizing luminosities are $6.3\times 10^{44}$~erg~s$^{-1}$ (for the unobscured SED) and $2.4\times 10^{44}$~erg~s$^{-1}$ (obscured SED).
 The  resulting obscured and unobscured SED models are shown in Fig.~\ref{fig:SED_comparison}.

\subsubsection{The Narrow Absorption Line Solution}

%
\begin{figure}[!tbp]
\centering
\resizebox{\hsize}{!}{\includegraphics[angle=0]{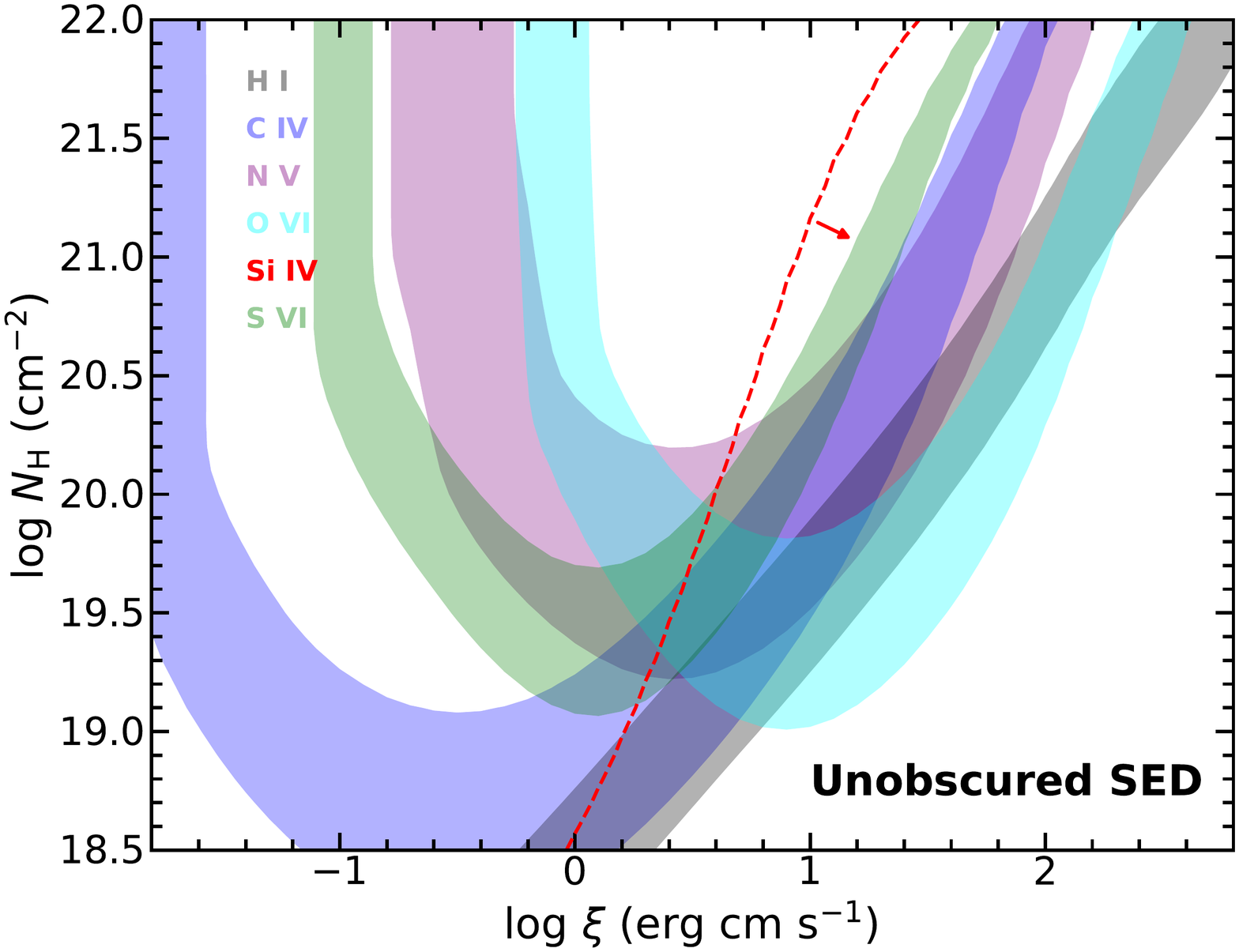}}
\resizebox{\hsize}{!}{\includegraphics[angle=0]{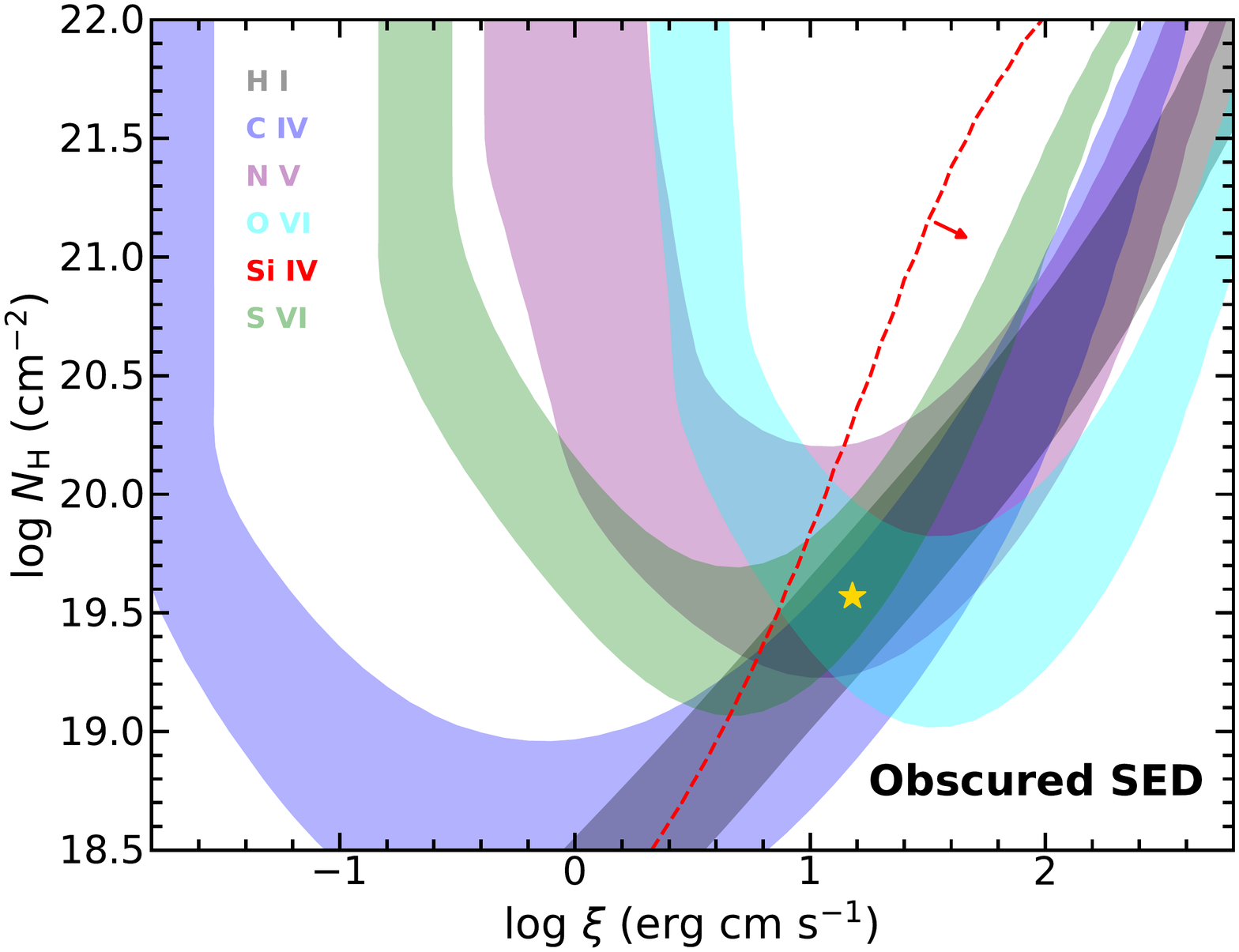}}
\caption{Photoionization model constraints on the narrow UV absorber in Mrk~817, calculated using the unobscured ({\it top panel}) and obscured ({\it bottom panel}) SEDs from the 2020-12-18 observation. The colored bands encompass the measured column density of each ion and its associated uncertainty. The dashed line for \ion{Si}{4} represents an upper-limit. Allowed photoionization solution would lie in a region where bands of all ions overlap. For the obscured SED case, there is a valid solution region at around ${\log N_{\rm H}\, ({\rm cm}^{-2}) = 19.5}$ and $\log \xi$ (erg~cm~s$^{-1}$)$= 1$ , indicated with a gold star.}
\label{narrow_sol}
\end{figure}

\begin{figure}[!tbp]
  \centering
   \resizebox{\hsize}{!}{\includegraphics[angle=0, width=\columnwidth]{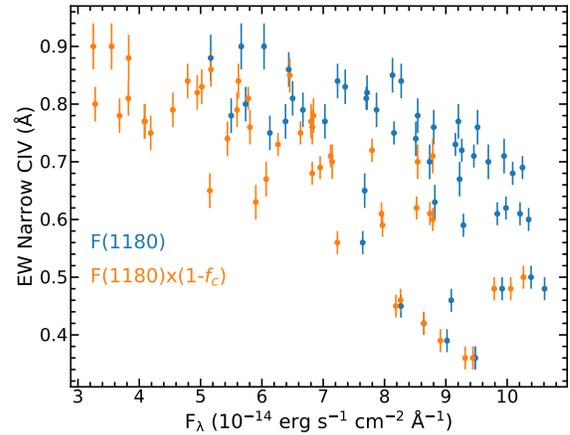}}
  \vskip 15pt
  \caption{Observed correlation between the equivalent width of narrow \ion{C}{4} $\lambda1548$ absorption and the continuum flux at 1180 \AA, F(1180) (blue points).
Orange points show how the correlation is improved by inferring an
obscured continuum flux by attenuating the observed flux by the transmission
of the deepest point in the \ion{C}{3}* absorption trough, $1 - f_c$.
}\label{fig_continuum_narrow}
\end{figure}

We used Cloudy v17.02 \citep{Ferl17} to determine the photoionization structure of the outflowing gas producing the narrow absorption lines in the HST spectra (Section~\ref{sec:uv_results}). We tried using the obscured and unobscured SEDs shown in Fig.~\ref{fig:SED_comparison}. The ionic column densities are computed with Cloudy over a grid of total column density $N_{\rm H}$ and ionization parameter $\xi$. The ionization parameter $\xi$ \citep{Kro81} is defined as ${\xi = {L}\, /\, {{n_{\rm{H}} r^2 }}}$ (in units of erg~cm~s$^{-1}$), where $L$ is the luminosity of the ionizing source over 1--1000 Ryd, $n_{\rm{H}}$ is the hydrogen density, and $r$ is the distance between the gas and the ionizing source. 

We set the elemental abundances to the proto-solar values of \citet{Lod09}. We also tried the default abundances of Cloudy \citep{ferland06}. The results from these two sets of abundances are similar.

 
 \begin{figure*}
\centering
\subfigure{\includegraphics[width=0.49\textwidth]{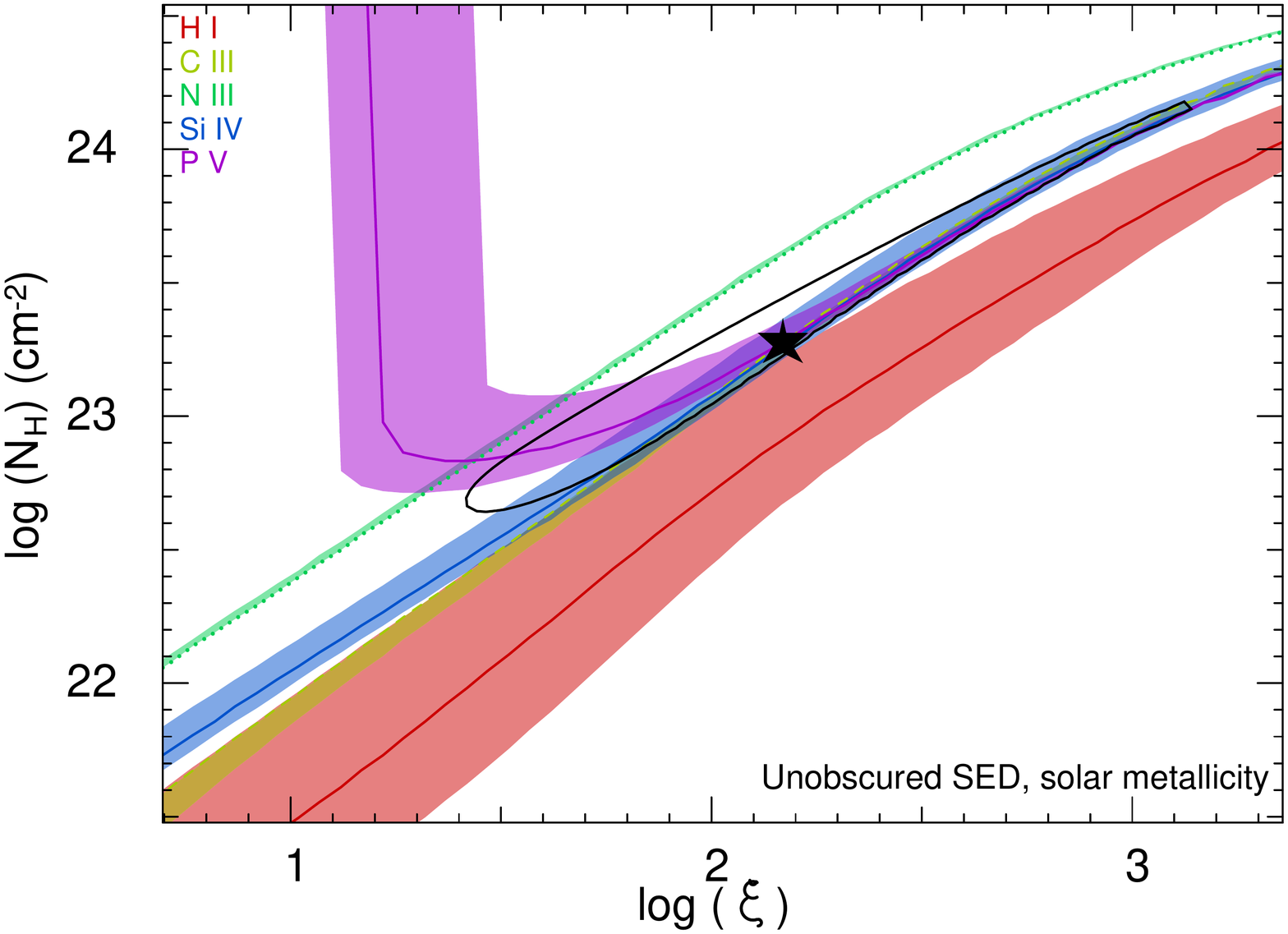}}
\subfigure{\includegraphics[width=0.49\textwidth]{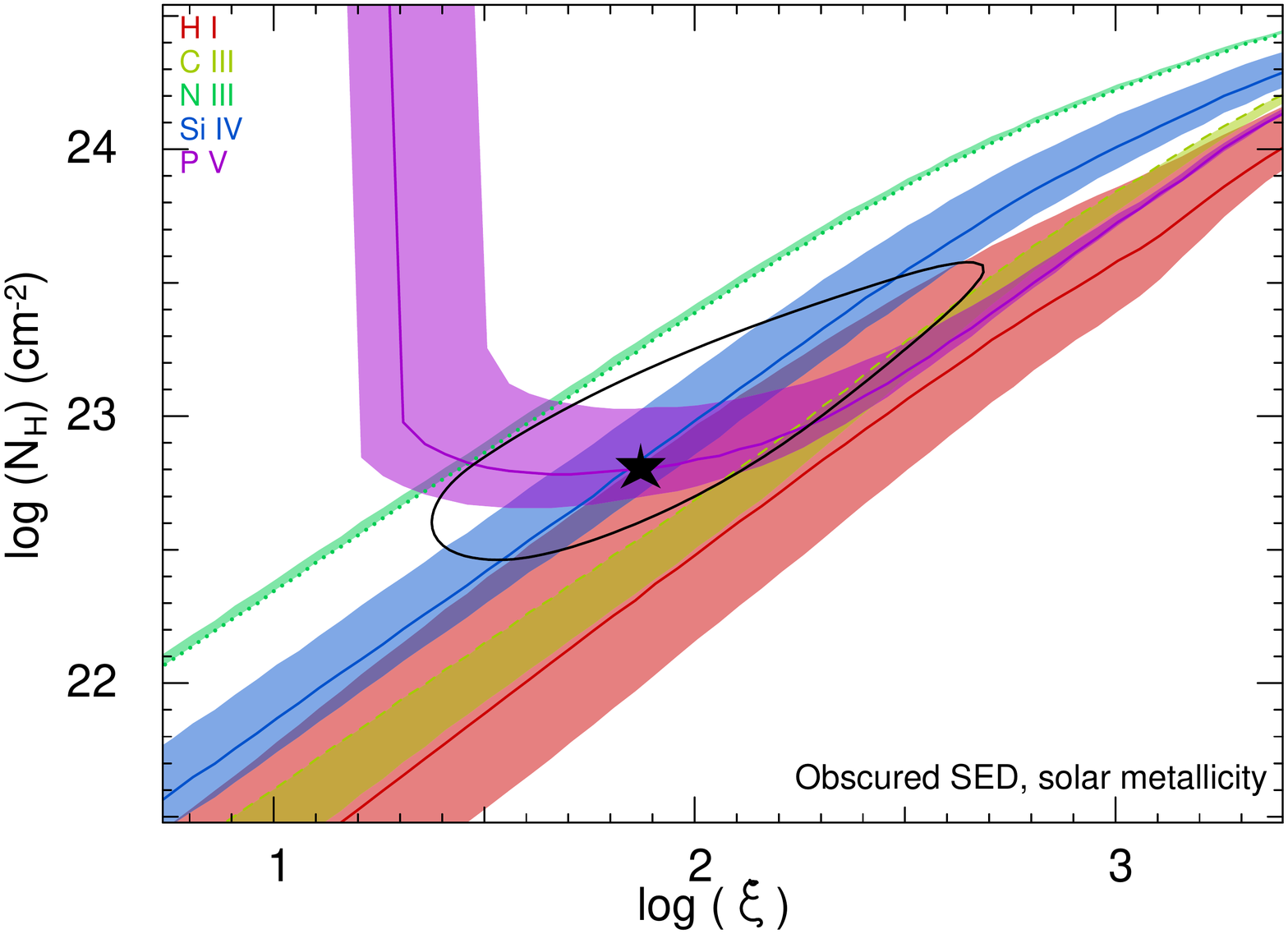}}
\caption{$\log{N_\mathrm{H}}$ vs $\log{\xi}$ for the unobscured ({\em left}) and obscured ({\em right}) SED models. The ions in the outflow system, represented by the colored bands, provide constraints on $N_\mathrm{H}$ and $\xi$ in the form of upper and lower limits, as well as measurement values, of measured column densities. Dotted lines are upper limits, dashed lines are lower limits, and solid lines are measurements. The uncertainty in the column density measurements are represented by the shaded bands. The solution found with $\chi^2$ minimization is represented by the black star, and the 1$\sigma$ is represented by a black ellipse.}
\label{fig:broad_photo}
\end{figure*}


Fig.~\ref{narrow_sol} shows the results for the narrow UV absorbers as observed on 2020-12-18. The calculations are done with the unobscured SED in the top panel, and with the obscured SED in the bottom panel. 
The column density constraints for each ion are shown by colored bands, and a good solution is one where the bands overlap. Encouragingly, Fig. \ref{narrow_sol} shows that there is no solution for the unobscured SED, while there is a solution for the obscured one at around ${\log N_{\rm H}\, ({\rm cm}^{-2}) = 19.5}$ and $\log \xi$ (erg~cm~s$^{-1}$)~$= 1$ (Fig. \ref{narrow_sol}-bottom panel). This suggests that the obscuring gas is located interior to the UV narrow-line absorber. 

This hypothesis is corroborated by the empirical result that the changes in the narrow absorption line equivalent width correlate better with the obscured continuum than the unobscured continuum (see Fig.~\ref{fig_continuum_narrow}). The obscured continuum is estimated by scaling the observed flux by the transmission fraction derived from the \ion{C}{3}* broad absorption trough. For the unobscured continuum, the Pearson correlation coefficient and $p$-value are $-0.61$ and $2.6\times10^{-6}$, respectively, while for the obscured continuum these are $-0.81$ and $6.7\times10^{-13}$.

\subsubsection{The Broad Absorption Line Solution}

Here we use the 2020-12-18 observations to determine the ionization structure of the gas producing the broad UV absorption troughs. First we determine the hydrogen number density ($n_\mathrm{H}$) by measuring the electron number density ($n_e$) using the \ion{C}{3}$^*$ absorption trough where in a fully ionized plasma $n_e\approx 1.2n_\mathrm{H}$. The ratio of column densities between each level of the \ion{C}{3}$^*$ multiplet and the \ion{C}{3} ground state is sensitive to $n_e$ and the temperature $T$ \citep[see][]{2005ApJ...631..741G,Arav15}. To measure $n_e$ we modeled the observed troughs with Gaussians (corresponding to the six \ion{C}{3}$^*$ transitions). We used the CHIANTI atomic database (CHIANTI 10.0) to model the ratios of the different excited states from $\log{n_e} = 3$ [cm$^{-3}]$ to $\log{n_e} = 13$ [cm$^{-3}]$. In order to determine the depth of the individual troughs, an estimate for the total column density of \ion{C}{3} was needed as well. 
The optical depths of the individual troughs are
\begin{eqnarray}
\tau_i = \frac{N_{\text{C III}} f_i \lambda_i}{3.8\times10^{14}~\text{cm}^{-2}}\times\frac{r_i}{\sqrt{2\pi}\sigma_{v}}
\end{eqnarray}
where $N_{\text{C III}}$ is the total column density of \ion{C}{3}, $f_i$ is the oscillator strength, $r_i$ is the ratio between the excited state and the total \ion{C}{3} column densities, and $\sigma_v$ is the Gaussian velocity width space determined by using the Lyman $\alpha$ absorption trough as a template. The troughs were modeled assuming apparent optical depth. We find that $\log{n_e}=10.5^{+0.8}_{-0.5}$ (in units of cm$^{-3}$) best fits the trough of \ion{C}{3}$^*$, with the errors determined through adjusting $n_e$ so that the $\chi^2$ increases by 1.

Similar to the previous section, using the obscured and unobscured SEDs in Fig.~\ref{fig:SED_comparison} and using the hydrogen number densities as computed above, we produced grids of Cloudy models with $\log{N_\mathrm{H}}$ (units cm$^{-2}$) ranging from 20 to 25.5, and $\log\xi$ ranging from 0.5 to 3.5, and search for the parameters that lead to ionic column densities closest to the measured values. Fig.~\ref{fig:broad_photo} shows the solution found with the unobscured SED ({\em left}) and obscured SED ({\em right}). The obscured solution meets all the ionic column density constraints and therefore is a viable physical model. The solution region (indicated by the black star in the ellipse) lies at roughly at ${\log N_{\rm H}\, ({\rm cm}^{-2}) = 22.8}$ and ${\log \xi = 1.9}$, which is formally higher than that found from X-ray spectral modelling (${\log \xi \sim 1}$, Section~\ref{sec:xray_results}). This is likely explained by differences in the photoionization codes/models (as described in \citealt{Mehdipour2016}).

Finally, for completeness, and motivated by previous work showing AGN with super-solar metallicity ouflows \citep{Gabel06,Arav07}, we created a grid of models with higher metallicity by applying the element ratios of \cite{2008A&A...478..335B}, using the recipe of   \cite{2020ApJS..247...41M} for five times solar metallicity ($Z_\odot$).  
Raising the metallicity  shifted the position of the \ion{H}{1} band relative to the other elements, and provides a slightly better solution compared to that found at solar metallicity.

\subsection{Early reverberation mapping results}
\label{sec:lags}

Next we present RM results of the first 97 days of the Mrk 817 campaign. The continuum RM results are given in Section~\ref{sec:continuum_rm} and the emission-line RM results in Section~\ref{sec:blr_rm}.

For both continuum and line reverberation, we calculate lags using the standard linear interpolation cross-correlation (ICCF) approach with uncertainties estimated from the flux randomization, random subset sampling technique \citep[as implemented by][]{peterson04}.  In this approach a large number of realizations (here we use $N = 10{,}000$) of the light curves are created through resampling of the data. In each realization the flux of each data point is randomized assuming a Gaussian distribution with a mean and standard deviation equal to the measured flux and its 1$\sigma$ uncertainty. In addition, the light curve is randomly sampled with replacement, meaning that some points are selected multiple times while others are not selected at all. Those that are selected multiple times have their errors weighted appropriately.  For each realization we measure the cross-correlation function (CCF) and its centroid value.  The lag is then taken as the median of the CCF centroid distribution, and its 1$\sigma$ uncertainty from the 16 and 84\% quantiles. 

\subsubsection{Continuum disk Reverberation Mapping}
\label{sec:continuum_rm}

We tested using several bands as the reference light curve. The HST 1180 \AA\ light curve has the highest variability amplitude, however, it resulted in a poor peak correlation coefficient ($R_{\rm max}$) with the longer wavelength optical bands. For instance, it gives an $R_{\rm max}$ of approximately 0.1 with the $z$ band.  The $g$ band light curve has the largest number of data points, but a significantly lower variability amplitude than the UV light curves. While it gave the best-constrained optical ground-based lags, the UV lags were significantly more poorly constrained. The Swift/\emph{UVW2}, on the other hand, has an excellent balance between the number of data points (nearly twice that of the HST 1180\AA\ light curve), high variability amplitude and good correlations with all wavebands. We therefore use this as the reference band against which we measure the lags.

\begin{figure}
    \centering
    \includegraphics[width=\columnwidth]{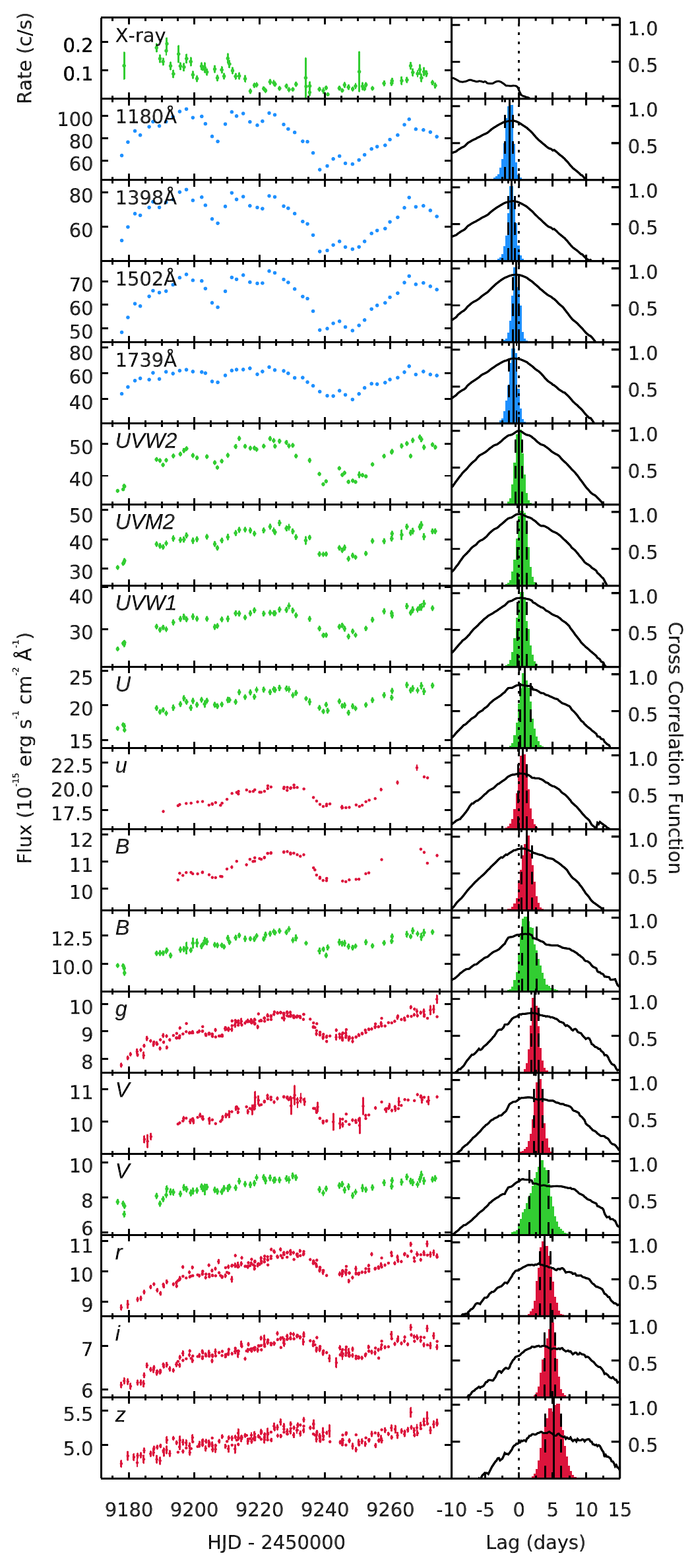}
    \caption{{\it Left:} Mrk 817 continuum light curves from HST (blue), Swift (green) and ground-based telescopes (red). All fluxes have units of $10^{-15}$~erg~s$^{-1}$~cm$^{-2}$~\AA$^{-1}$, aside from the {\it Swift} 0.3 -- 10 keV X-ray light curve which is shown as a count rate. {\it Right:} Solid black lines show the cross-correlation function with respect to the Swift \emph{UVW2} band. Colored histograms show the ICCF centroid lag distributions, while the vertical solid and dashed lines give the lag and 1$\sigma$ uncertainty range.}
    \label{fig:contlc}
\end{figure}

\begin{figure}
    \centering
    \includegraphics[width=\columnwidth]{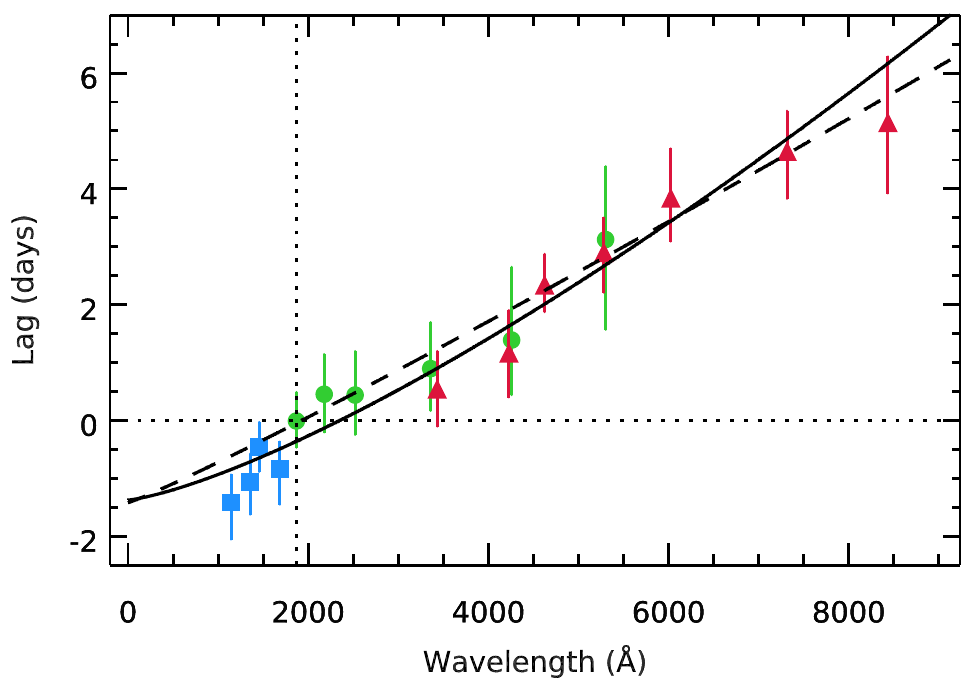}
    \caption{Continuum lags (rest-frame) calculated with respect to the {\it Swift/UVW2} band. Data from {\it HST}, {\it Swift} and ground-based telescopes are shown as blue squares, green circles, and red triangles respectively. The solid line shows the best-fitting $\tau\propto\lambda^{4/3}$ relation. The dashed line is the best-fitting model using the relations in \citet{kammoun21a}, with a maximal black hole spin.}
    \label{fig:contlags}
\end{figure}

The resulting rest-frame lags are given in Table~\ref{table:contlags}.  
We also give the fractional variability amplitude, $F_{\rm var}$ \citep{vaughan03}, which is largest in the UV and decreases with wavelength, and the maximum correlation coefficient $R_{\rm max}$.  The right-hand panels of Fig.~\ref{fig:contlc} show the CCFs (solid lines) and the CCF centroid distributions, while Fig.~\ref{fig:contlags} shows the lags as a function of wavelength.  The lags increase with wavelength, approximately following $\tau \propto \lambda^{4/3}$, as expected for a standard Shakura \& Sunyaev thin disk \citep{Cackett2007}.  We fit $\tau = \tau_0 \left[ \left(\lambda/\lambda_0\right)^\beta - y_0 \right]$, with $\lambda_0 = 1869$ \AA\ (the rest-frame wavelength of the {\it UVW2} band), and $\beta = 4/3$, and where $y_0$ allows the fit to be non-zero at $
\lambda_0$.  This gives a best-fitting value of $\tau_0 = 1.01\pm0.09$ days.

In the disk reprocessing scenario the lags should scale with black hole mass, $M$, and Eddington ratio, $\dot{m}_{\rm E}$ following $\tau \propto M^{2/3} \dot{m}_{\rm E}^{1/3}$.  Comparing to NGC~5548 \citep{fausnaugh16} we assume $M_{\rm NGC 5548} = 5\times10^7$~M$_\odot$, and $\dot{m}_{\rm E, NGC 5548} = 0.05$ and for Mrk 817 $M_{\rm Mrk 817} = 3.85\times10^7$~M$_\odot$, and $\dot{m}_{\rm E, Mrk 817} = 0.2$.  Given these values we expect $\tau_{0, \rm NGC 5548}/\tau_{0,\rm Mrk 817} = 0.75$.  Adjusting the best-fitting $\tau\propto\lambda^{4/3}$ to the NGC~5548 continuum lags to have the same reference wavelength as we use here we get $\tau_{0,\rm NGC 5548} = 0.64$ days. Therefore, the observed ratio of  $\tau_{0, \rm NGC 5548}/\tau_{0,\rm Mrk 817} = 0.63$ is consistent with the expected mass and Eddington ratio scaling given the uncertainties.

Recent analytical models for accretion disk lags have been developed using transfer functions calculated from general-relativistic ray-tracing simulations \citep{kammoun21a, kammoun21b}. We fit these models assuming both a non-spinning ($a=0$) and a maximally spinning ($a=0.998$) black hole. We fix the black hole mass at $M = 3.85\times10^7$~M$_\odot$, and use the unabsorbed 2 -- 10 keV X-ray flux from the XMM observations of $8.5\times10^{-12}$~erg~s$^{-1}$~cm$^{-2}$.  We leave the mass accretion rate and the height of the X-ray source, $h$, as free parameters in the fit, but, following \citet{kammoun21b} we constrain $h$ to be between 2.5 and 100 $R_G$.  Both the $a=0$ and $a=0.998$ models fit the data equally well, and we find that the height of the X-ray source is unconstrained. We get best-fitting mass accretion rates of $\dot{m}_{\rm Edd} = 0.06_{-0.02}^{+0.07}$ and $\dot{m}_{\rm Edd} = 0.32_{-0.12}^{+0.39}$ for $a=0$ and $a=0.998$ respectively. The best-fitting $a=0.998$ model is shown as a dashed line in Fig.~\ref{fig:contlags}.

All previous intensive campaigns that utilized Swift and ground-based monitoring \citep{edelson19,cackett18,vincentelli21,Hernandez2020} have found significant (typically a factor of $\sim$2) excess lags in u/U bands, relative to the adjacent bands or to the fits. By contrast, the lags presented here from the first $\sim$1/3 of the campaign show no evidence of excess $u/U$ band lags. In the one source, NGC 4593, where spectroscopic observations covering this wavelength range were available, \citet{cackett18} showed that this $u/U$ band excess was, in fact, a broad excess leading up to the Balmer jump.  This has been associated with lags from the diffuse continuum arising in the BLR gas \citep{koristagoad01,koristagoad19,lawther18, chelouche19,netzer20}. We will continue to monitor the $u/U$ band lags to better understand under what conditions this feature is or is not present.

Despite the strong and variable absorption observed in the X-rays and the UV absorption lines and the lack of an X-ray/UV correlation, the UV and optical continuum variability and continuum lags look very similar to what would be expected.  Either the disk sees a different source of irradiating photons than the X-rays we observe (e.g., the X-ray absorber is not located between the X-ray and UV/optical continuum region  or does not block all lines of sight between those regions), or a different mechanism drives variability in the disk \citep[e.g.\ corona-heated accretion disk reprocessing; ][]{sun20a,sun20b}.

\begin{deluxetable}{lcccC}
\label{table:contlags}
\tablewidth{0pt}
\tablecaption{Continuum lags and light curve properties}
\tablehead{
\colhead{Filter} &
\colhead{Telescope} &
\colhead{$F_{\rm var}$}  &
\colhead{$R_{\rm max}$} &
\colhead{Lag} \\
 & & & & \colhead{(days)}}
\startdata
1180\AA & HST & 0.177 & 0.80 & -1.42^{+0.48}_{-0.64} \\
1398\AA & HST & 0.155 & 0.81 & -1.06^{+0.46}_{-0.57}\\
1502\AA & HST & 0.119 & 0.92 &   -0.46\pm0.43\\
1739\AA & HST & 0.121 & 0.88 & -0.84^{+0.48}_{-0.61}\\
{\it UVW2} (1928\AA) & Swift & 0.103 & 1.00 & 0.00\pm0.50\\
{\it UVM2} (2246\AA) & Swift & 0.090 & 0.97 & 0.45^{+0.69}_{-0.66} \\
{\it UVW1} (2600\AA) & Swift & 0.078 & 0.94  & 0.44^{+0.76}_{-0.69}\\
{\it U} (3465\AA) & Swift & 0.071 & 0.86  &  0.90^{+0.81}_{-0.73}\\
{\it u}  (3540\AA) & Ground & 0.051 & 0.76 & 0.55\pm0.65 \\
{\it B}  (4361\AA) & Ground & 0.036 & 0.84 & 1.17^{+0.74}_{-0.77}\\
{\it B} (4392\AA) & Swift & 0.065 & 0.78  &  1.39^{+1.27}_{-0.95} \\
{\it g} (4770\AA) & Ground & 0.043 & 0.82 & 2.34^{+0.54}_{-0.48}\\
{\it V} (5448\AA) & Ground & 0.035 & 0.77 & 2.91^{+0.60}_{-0.70} \\
{\it V} (5468\AA) & Swift & 0.048 & 0.76  & 3.12^{+1.27}_{-1.56}\\
{\it r} (6215\AA) & Ground & 0.036 & 0.72 & 3.84^{+0.86}_{-0.76}\\
{\it i}	(7545\AA) & Ground & 0.037 & 0.70 & 4.65^{+0.70}_{-0.83}\\
{\it z}	(8700\AA) & Ground & 0.025 & 0.63 & 5.15^{+1.14}_{-1.23}\\
\enddata
\end{deluxetable}

\subsubsection{Broad-line region reverberation mapping}
\label{sec:blr_rm}

\begin{figure*}
    \centering
    \includegraphics[width=0.8\textwidth]{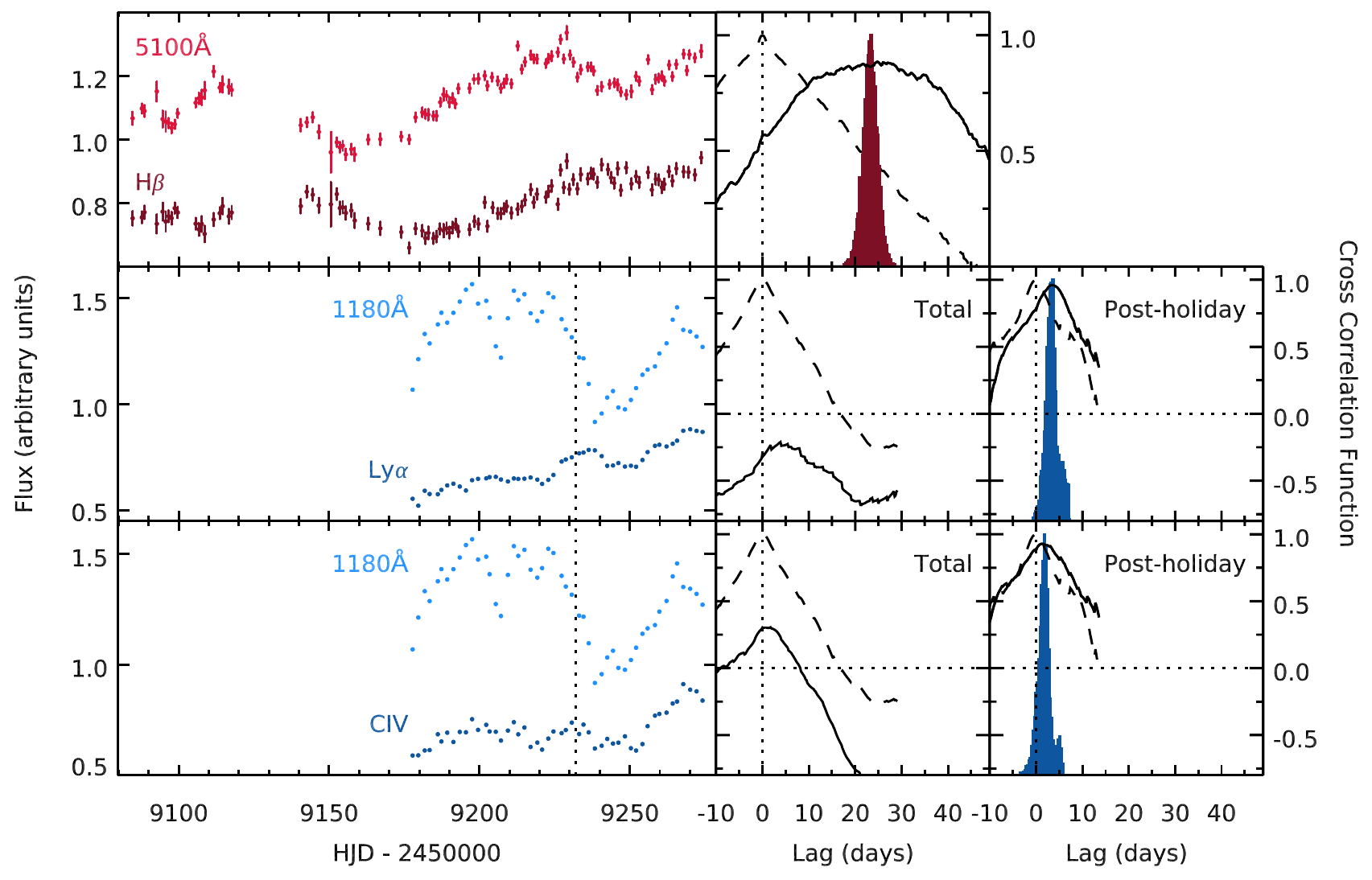}
    \caption{The broad line region lags from the AGN STORM 2 campaign to date. ({\em Top:}) The 5100 \AA\ continuum (red) and the \hb\ light curve (dark red). The panel on the right shows the cross correlation function (CCF; the solid black line) between these two bands, while the dashed line shows the auto correlation function (ACF) of the continuum. The colored histogram is the ICCF centroid lag distributions, resulting in an \hb\ lag of $23.2\pm1.6$ days (rest frame) behind the continuum. ({\em Middle, Bottom:}) The 1180 \AA\ continuum (blue) and the Ly$\alpha$ light curve (dark blue; {\em middle}) and \ion{C}{4} ({\em bottom}). The panels immediately to the right of the light curves show the CCF (solid line) and ACF (dashed line) of the entire campaign. We notice from the light curves that the correlation is much greater at HJD$>2459232$. We show the CCF and ACF and ICCF lag distribution for this `post-holiday' period in the far right panels. The resulting in time lags are $\tau_{\lya}=3.3^{+1.5}_{-1.2}$~days and $\tau_\mathrm{C~IV}=1.8^{+1.2}_{-1.3}$~days. We emphasize that these BLR lags are calculated from a subset of our data and will be updated as the campaign continues.} 
    \label{fig:blrlags}
\end{figure*}

The broad-line region reverberation mapping traces larger (centi-parsec) scales around the black hole. In order to probe these larger scales, the optical spectroscopy campaign began before the HST/Swift/NICER campaign. In Fig.~\ref{fig:blrlags}-{\em top}, we show the $5100$\AA\ continuum light curve (red) and the \hb\ light curve (dark red). The panel on the right shows the results of the ICCF analysis, searching for correlations over $-10$ to $50$ days (1/3 of the duration of the light curve). The peak of the cross-correlation coefficient is high ($R_{\mathrm{max}}$=0.89), and the time lag is measured to be $23.2\pm1.6$ days (rest frame). The time lags were also measured using {\sc Javelin} \citep{zu2011}, and the time lag was consistent ($22.0^{+2.0}_{-1.4}$ days). This is consistent with previous 5100\AA\ vs. \hb\ lags measured for this source from previous campaigns \citep{zu2011}.

Thanks to the dedicated \HST\ campaign, we can also measure the UV broad line lags. The light curves for \lya\ and \ion{C}{4} are shown in the middle and bottom panels of Fig.~\ref{fig:blrlags}, and ACFs, CCFs and ICCF lag distributions are shown in the panels immediately to the right (labeled `total'). Again, we search for correlations over a time-frame that is 1/3 the duration of the light curve. Calculating the CCF for the total campaign thus far, we found little to no correlation between the UV lines and the 1180 \AA\ continuum. This is indicated by the solid black line cross correlation functions that peak at $R_{\mathrm{max}}=-0.2$ and $R_{\mathrm{max}}=0.3$ for \lya\ and \civ\, respectively. These correlations are so low that we cannot determine a lag. 

However, we notice a change, starting at HJD=2459232, when the UV lines {\em do} begin to correlate with the continuum. The panels on the far right show the ACFs, CCFs and ICCF lag distributions from HJD=2459232 and beyond (labeled 'post-holiday'). The post-holiday light curves show $R_{\mathrm{max}}$ of nearly unity (0.96 for \lya\ and 0.93 for \ion{C}{4}), and we measure the lags to be $\tau_{\mathrm{\lya}}=3.3^{+1.5}_{-1.2}$~days and $\tau_{\mathrm{C~IV}}=1.8^{+1.2}_{-1.3}$~days (rest frame). We emphasize that this period of correlated variability spans only 43~days, a small subset of our total campaign, and so the lags should be regarded as preliminary. Further measurements including velocity-resolved reverberation mapping and updated black hole mass estimates will be carried out using data from the full duration of the campaign.

\section{Discussion}

In the canonical reverberation mapping paradigm, the variability we observe is thought to be driven by rapid variability in the X-ray corona. This driving light curve irradiates other gas flows, and those light echoes across different wavebands allow us to map out scales from the inner edge of the accretion disk (e.g., \citealt{zoghbi10,kara16}), to the outer accretion disk (e.g., \citealt{shappee14,Edelson2015, cackett18}) to the broad line region (\citealt{peterson04,bentz13}). While much of this paradigm has been successful in explaining the variability and time delays observed in dozens of Seyfert 1 AGN, there are important exceptions, that fundamentally can teach us more about the complex nature and dynamics of gas flows around supermassive black holes. For instance, variable line of sight obscuration (e.g., \citealt{zoghbi19,Dehghanian2019}), variability due to mass accretion rate fluctuations in the disk (e.g., \citealt{mchardy14}) or variability in the geometry of the corona (e.g., \citealt{Alston2020}) are not included in the simplest reverberation paradigm, but indeed, are likely important for understanding the entire system.

Understanding when and why the simple paradigm works or does not work requires large spectroscopic and photometric multi-wavelength efforts spanning a range of timescales that allow us to put the AGN's variability into context. This is the goal of the AGN STORM 2 campaign, where we have consolidated efforts from across many AGN fields in order to get a complete picture of the Seyfert galaxy Mrk 817. When the campaign began in November 2020, we discovered Mrk~817 in a never-before-seen obscured state, with a depressed soft X-ray flux (Fig.~\ref{fig:x-ray}) and broad UV absorption troughs (Fig.~\ref{fig_cosfull}) suggesting that an ionized, dust-free absorber located at the inner broad line region partially obscures our line of sight to the central engine.

The obscuration strongly affects the soft X-ray flux (Fig.~\ref{fig:evolution}), which, naively, one might think explains the lack of correlation between X-rays and longer wavelengths (Fig.~\ref{fig:contlc}). However, earlier Swift monitoring of Mrk~817 (when the source was not obscured) also show little correlation between X-rays and longer wavelengths \citep{morales19}. Moreover, this lack of correlation is commonly seen in other AGN that do not exhibit time-variable obscuration \citep{buisson17, edelson19}, and thus likely points to additional complexities in the coupling between the accretion disk and the corona.  Despite a lack of correlation between the observed driving light curve (X-rays) and longer wavelengths, we still observe UV and optical continuum lags as expected from a thin disk being irradiated and heated by a compact ionizing source (Fig.~\ref{fig:contlags}). In future work with the complete dataset, we will also test models including a diffuse continuum from the broad line region that may contribute to the continuum lags.

The standard reverberation mapping paradigm also explains the 23-day time lag between the 5100 \AA~continuum light curve and the \hb\ light curve as due to the light travel time between the accretion disk and BLR (Fig.~\ref{fig:blrlags}-{\em top}). This lag agrees  well with the radius-luminosity relation \citep{bentz13}. That said, the UV lines like \lya\ and \ion{C}{4} are more complicated, as the first 55 days (before HJD 2459232) show a much weaker correlation between continuum and the \civ\ broad emission line than later epochs (Fig.~\ref{fig:blrlags}-{\em bottom}).  The $\sim 2-3$ day lags of the UV lines are much shorter than the \hb\ lags, and in fact, are comparable to the lags between UV and the NIR continuum, which may suggest that the UV broad line region and the outer accretion disk are co-spatial. Future work modelling the reverberation lags will put better constraints on the geometry and dynamics of the accretion disk and broad line region, but for reference, 1 light-day corresponds to a distance of 455~$GM/c^2$ in units of gravitational radii.  

A de-coupling of the UV continuum and broad emission lines was seen also in the AGN~STORM~1 campaign in NGC~5548 for a period of 60--70 days \citep{Goad2016}. Because the anomalous period was more exaggerated in the higher ionization lines (like \civ\ and \siIV) than in lower ionization lines like \lya, \citet{Goad2016} suggested that the de-coupling was due to a depletion of ionizing photons above $E>56$~eV, relative to those near 13.6 eV. This could either be caused by intrinsic changes in the X-ray corona or due to line-of-sight obscuration. In the case of Mrk~817, the UV holiday occurs at the beginning of the campaign, as X-ray emission was becomes more obscured, and the strength of the broad UV absorption troughs increases (Fig.~\ref{fig:evolution}). Then, as the line-of-sight obscuration decreases, we measure coherent lags between the UV continuum and broad emission lines. This may suggest that obscuration effects are responsible for the de-coupling of continuum and UV broad lines. As the campaign continues, we will be able to better measure time lags associate with the various broad emission lines.

From the first third of the AGN~STORM~2 Campaign targeting Mrk~817, we have identified and characterized a new ionized obscurer. With an ionizing luminosity of $2.4\times 10^{44}$~erg~s$^{-1}$ for the obscured SED, our measurement of an electron number density of log $n_e = 10.5$, and the best fit ionization parameter of $\log \xi=1.9$, the UV absorbing gas in the obscurer lies at a radius of $R=3.7^{+0.9}_{-1.1}$~light days. This location is consistent with the inner BLR, which, from our preliminary reverberation lag analysis, we suggest is $\sim 3$ light-days ($\sim 1500~GM/c^{2}$) from the black hole. Continuum reverberation lags also suggest that the accretion disk extends out this far (and beyond), and so we suggest that the ionized obscurer is associated with an accretion disk wind.

Similar accretion disk wind models have been envoked to explain the changes to the broadband SED \citep{mehdipour16}, high ionization absorption lines \citep{Kriss2019} and the broad line region holiday in NGC~5548 \citep{Dehghanian2019,Dehghanian2020}. In such a model, the equatorial accretion disk wind is densest at the base and more diffuse at larger scale heights. Some numerical simulations of disk winds, especially of radiation driven disk winds (e.g., \citealt{murray95,proga00,proga07}), also show that the wind base can be a crucial site of line emission and absorption, and can affect broad line reverberation lags
(e.g., \citealt{chiang96,proga04,kashi13,waters16,giustini19}). Comparisons between theoretical models and observations will be presented in future work.

In the case of Mrk~817, it is possible that at the very beginning of the campaign, a lower density, larger scale height wind component is present, and accounts for modest X-ray and UV light-of-sight obscuration. The base of the wind, however, is dense enough to interfere with UV BLR irradiation. 
As the campaign continues through HJD~2459232, the dense wind flows outwards, and the line-of-sight obscuration increases (as shown in Fig.~\ref{fig:evolution}). With the ejection of the densest part of the wind, the BLR can again `see' the accretion disk, and the UV BLR reverberation ensues. As the gas reservoir of the wind is depleted, the line-of-sight obscuration decreases from HJD~2459232 onwards.    
This unveiling of the central source will have important implications for future observations later on in the campaign, allowing us to put this model to the test.

\section{Summary and Conclusions}

To summarize, our major findings from the first third of the AGN~STORM~2 campaign of Mrk~817 (November 2020--March 2021) are as follows: 
\begin{itemize}
    \item Compared to archival observations from 2019, the soft X-ray flux dropped by a factor of $\sim 10$. This variability suggests the presence of a partially covering, ionized obscurer.
    \item The UV continuum did not drop relative to archival observations, but the \HST\ spectra revealed new blueshifted absorption lines. 
    \item Analysis of the narrow \ion{N}{5} doublet and the broad \ion{O}{6} doublet reveal that the broad and narrow absorbers cover only the continuum emission, and therefore likely originate at the inner BLR or within.
    \item The photoionization solution for the narrow and broad absorption UV lines is most self-consistently explained if the ionizing SED is obscured. This is supported by the fact that the narrow absorption lines correlate better with a continuum that is obscured by the broad absorption component.
    \item Disk continuum lags: The UV/optical continuum lags (on the order of days) are consistent with a centrally illuminated Shakura-Sunyaev thin accretion disk, which may suggest that the absorber may be beyond the inner accretion disk. 
    \item Optical broad line region lags: The \hb\ emission line lags the optical continuum by 23~days, similar to previous \hb\ reverberation mapping campaigns of this source.
    \item UV broad line region lags: The first 55~days of the campaign showed little correlation between the UV continuum and UV broad lines like \lya\ and \civ\, but from the next 42 days (as the obscuration appears to decrease), we measure UV BLR lags of 2-3~days.
\end{itemize}

We continue to monitor Mrk~817 across the electromagnetic spectrum. At the time of writing, the X-ray flux is increasing and the broad UV absorption lines are decreasing in equivalent width, perhaps suggesting that we are entering a new, unobscured phase of the campaign.

\section*{Acknowledgements}

The AGN~STORM~2 collaboration thanks \xmm\ Project Scientist, Norbert Schartel, for approving our ToO request. Special thanks as well to the \xmm\ Science Operations Center Coordinator, Ignacio de la Calle, and \nustar\ Science Operations Manager, Karl Forster, for help coordinating with multiple facilities.
Support for Hubble Space Telescope program GO-16196 was provided by NASA through a grant from the Space Telescope Science Institute, which is operated by the Association of Universities for Research in Astronomy, Inc., under NASA contract NAS5-26555.
 We are grateful to the dedication of the Institute staff who worked hard to review and implement this program. We particularly thank the Program Coordinator, W. Januszewski, who is making sure the intensive monitoring schedule and coordination with other facilities continues successfully.
 This work made use of data supplied by the UK Swift Science Data Centre at the University of Leicester.

This work makes use of observations from the Las Cumbres Observatory global telescope network. 
The Liverpool Telescope is operated on the island of La Palma by Liverpool John Moores University in the Spanish Observatorio del Roque de los Muchachos of the Instituto de Astrofisica de Canarias with financial support from the UK Science and Technology Facilities Council.
We thank WIRO engineers Conrad Vogel and Andrew Hudson for their invaluable  assistance. A major upgrade of the Kast spectrograph on the Shane 3~m telescope at Lick 
Observatory was made possible through generous gifts from the Heising-Simons 
Foundation as well as William and Marina Kast. Research at Lick Observatory is 
partially supported by a generous gift from Google.

This work makes use of observations collected at the Centro Astronómico Hispanoen Andalucía (CAHA) at Calar Alto, operated jointly by the Andalusian Universities and the Instituto de Astrofísica de Andalucía (CSIC). Funding for the Lijiang 2.4m telescope has been provided by Chinese Academy of Sciences (CAS) and the People's Government of Yunnan Province.

E.M.C. and J.A.M. gratefully acknowledge support from NSF grant AST-1909199. Research at UC Irvine was supported by NSF grant AST-1907290. H.L. acknowledges a Daphne Jackson Fellowship sponsored by the Science and Technology Facilities Council (STFC), UK. J.A.J.M. and M.J.W. acknowledge support from STFC grants ST/P000541/1 and ST/T000244/1. G.J.F. and M.D. acknowledge support by NSF (1816537, 1910687), NASA (ATP 17-ATP17-0141, 19-ATP19-0188), and STScI (HST-AR- 15018 and HST-GO-16196.003-A). Y.H. acknowledge support from NASA grants HST-GO-15650 and HST-GO-16-196. D.I. and L. \v C. P. acknowledge funding provided by the Astronomical Observatory Belgrade (the contract 451-03-68/2020-14/200002), University of Belgrade - Faculty of Mathematics (the contract 451-03-68/2020-14/200002)
through the grants by the Ministry of Education, Science, and Technological Development of the Republic of Serbia. D.I. acknowledges the support of the Alexander von Humboldt Foundation.  M.V. gratefully acknowledges financial support from the Independent Research Fund Denmark via grant number DFF 8021-00130. J.M.W. acknowledges financial support by the National Science Foundation of China (NSFC) through grants NSFC-11833008 and -11991054. P.D. acknowledges financial support from NSFC grants NSFC-12022301, -11873048, and -11991051, and from the Strategic Priority Research Program of the CAS (XDB23010400). C.H. acknowledges financial support from the grant NSFC-11773029 and from the National Key R\&D Program of China (2016YFA0400701). Y.R.L. acknowledges financial support from the grant NSFC-11922304 and from the Youth Innovation Promotion Association CAS. J.V.H.S. and K.H. acknowledge support from STFC grant ST/R000824/1. P.B.H. is supported by NSERC. Support for A.V.F.'s group at U.C.Berkeley is provided by the TABASGO Foundation, 
the Christopher R. Redlich Fund, and the Miller Institute for Basic Research in Science
(A.V.F. is a Senior Miller Fellow). MCB gratefully acknowledges support from the NSF through grant AST-2009230.
D.K. acknowledges support from the Czech Science Foundation project No. 19-05599Y.

\facilities{HST (COS, STIS), XMM, NuSTAR, NICER, Swift, LCO, Liverpool:2m, Wise Observatory, Zowada, CAO:2.2m, FTN, Shane, YAO:2.4m, WIRO, ARC, Gemini:Gillett, IRTF}

\appendix

\section{Summary of UV line characteristics}

To characterize the absorption lines more quantitatively,
we can empirically measure the centroid of an absorption feature to
obtain the mean outflow velocity, its full-width at half maximum (FWHM), and
its Equivalent Width (EW; integrated normalized flux).
Table \ref{tab_abs_lines} summarizes these properties for the narrow and broad
absorption lines in Mrk 817 for Visit 3N.

Table~\ref{tab_columns} shows the column densities of the UV narrow and broad absorption lines that were used for the photoionization modelling in Section~\ref{sec:photoionization}.

Details of the \ion{H}{1} column densities can be found in Table~\ref{tab_h1}.

\begin{deluxetable*}{lccccc}
\tablecaption{Absorption Line Properties in Mrk 817 Visit 3N}\label{tab_abs_lines}
\tablehead{
\colhead{Line} & \colhead{$\lambda_{rest}$} & \colhead{$\rm v_{out}$\tablenotemark{a}} & \colhead{FWHM\tablenotemark{b}} & \colhead{EW\tablenotemark{c}} & \colhead{$\rm f_c$\tablenotemark{d}} \\
\colhead{} & \colhead{$\rm km~s^{-1}$} & \colhead{$\rm  km~s^{-1}$} & \colhead{\AA}
}
\startdata
\multicolumn{6}{c}{Narrow Absorption Lines} \\
\hline
\ion{S}{6}  &  933.38 & $-3722$ &  79 & $0.20 \pm 0.07$ &  0.70 \\
\ion{S}{6}  &  944.52 & $-3722$ &  79 & $0.13 \pm 0.07$ &  0.46 \\
\ion{O}{6}  & 1031.93 & $-3709$ & 219 & $0.84 \pm 0.13$ &  1.00 \\
\ion{O}{6}  & 1037.62 & $-3709$ & 219 & $0.84 \pm 0.13$ &  1.00 \\
Ly$\alpha$  & 1215.67 & $-3709$ & 219  & $0.56 \pm 0.02$ &  0.58 \\
\ion{N}{5}  & 1238.82 & $-3726$ & 161 & $0.66 \pm 0.04$ &  0.91 \\
\ion{N}{5}  & 1242.80 & $-3726$ & 161 & $0.66 \pm 0.04$ &  0.90 \\
\ion{C}{4}  & 1548.19 & $-3722$ & 173 & $0.69 \pm 0.05$ &  0.71 \\
\ion{C}{4}  & 1550.77 & $-3722$ & 173 & $0.56 \pm 0.05$ &  0.58 \\
\hline
\multicolumn{6}{c}{Broad Absorption Lines} \\
\hline
\ion{S}{6}  &  933.38 & $-5591$ & 922 & $1.78 \pm 0.37$ &  0.57 \\
\ion{S}{6}  &  944.52 & $-5591$ & 922 & $1.70 \pm 0.37$ &  0.55 \\
\ion{N}{3}  &  990.68 & $-5510$ & 1165 & $0.36 \pm 0.406$ &  0.087 \\
\ion{C}{3}  &  977.02 & $-5510$ & 1165 & $0.51 \pm 0.344$ &  0.124 \\
\ion{O}{6}  & 1031.93 & $-5510$ & 1165 & $3.46 \pm 0.26$ &  0.79 \\
\ion{O}{6}  & 1037.62 & $-5510$ & 1165 & $3.39 \pm 0.26$ &  0.77 \\
\ion{P}{5}  & 1117.98 & $-5510$ & 1165 & $0.56 \pm 0.03$ &  0.12 \\
\ion{P}{5}  & 1128.01 & $-5510$ & 1165 & $0.55 \pm 0.03$ &  0.12 \\
\ion{C}{3}* & 1175.74 & $-5510$ & 1165 & $0.66 \pm 0.022$ &  0.133 \\
Ly$\alpha$  & 1215.67 & $-5561$ & 1165 & $3.03 \pm 0.027$ &  0.59 \\
\ion{N}{5}  & 1238.82 & $-5510$ & 1165 & $1.57 \pm 0.12$ &  0.30 \\
\ion{N}{5}  & 1242.80 & $-5510$ & 1165 & $1.50 \pm 0.12$ &  0.28 \\
\ion{Si}{4} & 1393.76 & $-5476$ & 911 & $0.81 \pm 0.05$ &  0.18 \\
\ion{Si}{4} & 1402.77 & $-5476$ & 911 & $0.71 \pm 0.05$ &  0.16 \\
\ion{C}{4}  & 1548.19 & $-5479$ & 1353 & $3.29 \pm 0.15$ &  0.44 \\
\ion{C}{4}  & 1550.77 & $-5479$ & 1353 & $2.96 \pm 0.15$ &  0.40 \\
\enddata
\tablenotetext{a}{Outflow velocities are relative to a systemic redshift of 0.0031455 \citep{Strauss88}. The uncertainty is dominated by systematic errors at 5 $\rm km~s^{-1}$.}
\tablenotetext{b}{Full width at half maximum}
\tablenotetext{c}{Equivalent width relative to the continuum flux.}
\tablenotetext{d}{Maximum possible covering fraction as measured at the deepest point of the absorption trough.}
\end{deluxetable*}

\begin{deluxetable*}{lccc}
\tablecaption{Absorption Line Column Densities in Mrk 817 Visit 3N}\label{tab_columns}
\tablehead{
\colhead{Ion} & \colhead{Best-fit $\rm log~N_{ion}$} & \colhead{$\rm log~N_{ion}$ Lower Limit} & \colhead{$\rm N_{ion}$ Upper Limit}\\
\colhead{} & \colhead{($\rm cm^{-2}$)} & \colhead{($\rm cm^{-2}$)} & \colhead{($\rm cm^{-2}$)}
}
\startdata
\multicolumn{4}{c}{Narrow Absorption Lines} \\
\hline
\ion{S}{6}  & 14.00 & 13.64 & 14.27 \\
\ion{O}{6}  & 15.58 & 15.33 & 16.13 \\
\ion{H}{1}  & 14.09 & 14.00 & 14.64 \\
\ion{N}{5}  & 14.79 & 14.69 & 14.84 \\
\ion{C}{4}  & 14.41 & 14.35 & 14.47 \\
\hline
\multicolumn{4}{c}{Broad Absorption Lines} \\
\hline
\ion{S}{6}  & 15.95 & 15.28 & 16.00\tablenotemark{a} \\
\ion{N}{3}  & 16.48 & 10.00 & 16.54 \\
\ion{C}{3}  & 16.04 & 10.00 & 16.10 \\
\ion{O}{6}  & 16.82 & 16.57 & 16.89\tablenotemark{a} \\
\ion{P}{5}  & 16.02 & 15.88 & 16.29 \\
\ion{C}{3}* & 15.91 & 15.90 & 15.91 \\
\ion{H}{1}  & 16.20 & 15.81 & 16.55 \\
\ion{N}{5}  & 16.68 & 15.00 & 16.98\tablenotemark{a,b} \\
\ion{Si}{4} & 15.81 & 15.64 & 16.06 \\
\ion{C}{4}  & 16.58 & 15.75 & 16.86\tablenotemark{a} \\
\enddata
\tablenotetext{a}{Upper limit could be much higher due to the high level of saturation.}
\tablenotetext{b}{Highly uncertain due to blending with the Ly$\alpha$ emission-line profile.}
\end{deluxetable*}

\begin{deluxetable*}{lccc}
\tablecaption{\ion{H}{1} Column Densities from Broad Lyman Lines in Mrk 817 Visit 3N}\label{tab_h1}
\tablehead{
\colhead{Lyman Line} & \colhead{Best-fit $\rm N_{HI}$} & \colhead{AOD $\rm N_{HI}$}\tablenotemark{a} & \colhead{$\rm N_{HI}$ Upper Limit}\\
\colhead{} & \colhead{($\rm 10^{16}~cm^{-2}$)} & \colhead{($\rm 10^{16}~cm^{-2}$)} & \colhead{($\rm 10^{16}~cm^{-2}$)}
}
\startdata
Ly$\alpha$ & 2.00 & 0.047 & 2.02 \\
Ly$\beta$  & 0.90 & 0.14  & 2.00 \\
Ly$\gamma$ & 1.40 & 0.36  & 3.51 \\
Ly$\delta$ & 1.60 & 0.60  & 4.89 \\
\enddata
\tablenotemark{a}{Apparent Optical Depth as calculated from a direct integration of the normalized line profile as described by \citet{Savage91}.}
\end{deluxetable*}


\bibliographystyle{apj}
\bibliography{ref.bib}

\end{document}